\documentclass[traditabstract,longauth]{aa} 
\usepackage{graphicx}
\usepackage{txfonts}
\usepackage{natbib}
\usepackage{longtable}

\newcommand{\HI}{{\rm H~{\sc i}}} 
\newcommand{\CIV}{{\rm C~{\sc iv}}} 
\newcommand{\CIII}{{\rm C~{\sc iii]}}} 
\newcommand{\AlIII}{{\rm Al~{\sc iii}}}
\newcommand{\SiIV}{{\rm Si~{\sc iv}}} 
\newcommand{\MgII}{{\rm Mg~{\sc ii}}}

\newcommand{\lya}{Lyman-$\alpha$}

\def\Sec#1{Section~\ref{s:#1}}
\def\App#1{Appendix~\ref{ap:#1}}

\def\Fig#1{Fig.~\ref{fig:#1}}
\def\Tab#1{Table~\ref{t:#1}}
\def\ion#1#2{{\rm #1~{\sc #2}}}
\defcitealias{hewett2010}{HW10}

\begin{document}
   \title{The Sloan Digital Sky Survey quasar catalog: tenth data release}

   \author{Isabelle P\^aris
          \inst{1}
          \and
          Patrick Petitjean
          \inst{2}
          \and
          \'Eric Aubourg
          \inst{3}  
          \and
          Nicholas P. Ross
          \inst{4}
          \and
          Adam D. Myers
          \inst{5,6}
          \and
          Alina Streblyanska
          \inst{7,8}
          \and
          Stephen Bailey
          \inst{4}
          \and
          Patrick B. Hall
          \inst{9}
          \and
          Michael A. Strauss
          \inst{10}
          \and
          Scott F. Anderson
          \inst{11}
          \and
          Dmitry Bizyaev
          \inst{12}
          \and
          Arnaud Borde
          \inst{13}
          \and
          J. Brinkmann
          \inst{12}
          \and
          Jo Bovy\thanks{Hubble fellow}
          \inst{14}
          \and
          William N. Brandt
          \inst{15,16}
          \and
          Howard Brewington
          \inst{12}
          \and
          Joel R. Brownstein
          \inst{17}
          \and
		  Benjamin A. Cook
		  \inst{10}
          \and
          Garrett Ebelke
          \inst{12}
          \and
          Xiaohui Fan
          \inst{18}
          \and
          Nurten Filiz Ak
          \inst{15,16,19}
          \and
          Hayley Finley
          \inst{2}
          \and
          Andreu Font-Ribera
          \inst{4,20}
          \and
          Jian Ge
          \inst{21}
          \and
          Fred Hamann
          \inst{21}
          \and
          Shirley Ho
          \inst{22}
          \and
          Linhua Jiang
          \inst{18}
          \and
          Karen Kinemuchi
          \inst{12}
          \and
          Elena  Malanushenko
          \inst{12}
          \and
          Viktor Malanushenko
          \inst{12}
          \and
          Moses Marchante
          \inst{12}
          \and
          Ian D. McGreer
          \inst{18}
          \and
          Richard G. McMahon
          \inst{23}
          \and
          Jordi Miralda-Escud\'e
          \inst{24,25}
          \and
          Demitri Muna
          \inst{26}
          \and
          Pasquier Noterdaeme
          \inst{2}
          \and
          Daniel Oravetz
          \inst{12}
          \and
          Nathalie Palanque-Delabrouille
          \inst{13}
          \and
          Kaike Pan
          \inst{12}
          \and
          Isma\"el Perez-Fournon
          \inst{7,8}
          \and
          Matthew Pieri
          \inst{27}
          \and
          Rog\'erio Riffel
          \inst{28,29}
          \and
          David J. Schlegel
          \inst{4}
          \and
          Donald P. Schneider
          \inst{15,16}
          \and
          Audrey Simmons
          \inst{12}
          \and
          Matteo Viel
          \inst{30,31}
          \and
          Benjamin A. Weaver
          \inst{32}
          \and
          W. Michael Wood-Vasey 
          \inst{33}
          \and
          Christophe Y\`eche
          \inst{13}
          \and
          Donald G. York
          \inst{34,35} 
 }
   \institute{Departamento de Astronom\'ia, Universidad de Chile, Casilla 36-D, Santiago, Chile
\email{paris@iap.fr}
         \and
            UPMC-CNRS, UMR7095, Institut d'Astrophysique de Paris, F-75014, Paris, France,              
         \and
         	APC, Astroparticule et Cosmologie, Uniiversit\'e Paris Diderot, CNRS/IN2P3, CEA/Irfu, Observatoire de Paris, Sorbonne Paris Cit\'e, 10, rue Alice Domon \& L\'eonie Duquet, 75205 Paris Cedex 13, France
		 \and
			Lawrence Berkeley National Laboratory, 1 Cyclotron Road, Berkeley, CA 94720, USA
         \and
         	Department of Physics and Astronomy, University of Wyoming, Laramie, WY 82071, USA
     	 \and
     	 	Max-Planck-Institut f\"ur Astronomie, K\"onigstuhl 17, D-69117 Heidelberg, Germany
     	 \and
     	 	 Instituto de Astrofisica de Canarias (IAC), E-38200 La Laguna, Tenerife, Spain
     	 \and
     	 	Universidad de La Laguna (ULL), Dept. Astrofisica, E-38206 La Laguna, Tenerife, Spain
     	 \and
     	 	Department of Physics and Astronomy, York University, Toronto, ON M3J1P3, Canada
      	 \and
     	 	Princeton University Observatory, Peyton Hall, Princeton, NJ 08544, USA
     	 \and
     	 	University of Washington, Dept. of Astronomy, Box 351580, Seattle, WA 98195, USA
     	 \and
     	 	Apache Point Observatory, P.O. Box 59, Sunspot, NM 88349-0059, USA
     	 \and
     	 	CEA, Centre de Saclay, Irfu/SPP, 91191, Gif-sur-Yvette, France
     	 \and
     	 	Institute for Advanced Study, Einstein Drive, Princeton, NJ 08540, USA
     	 \and
     	 	Department of Astronomy and Astrophysics, The Pennsylvania State University, University Park, PA 16802, USA
     	 \and
     	 	Institute for Gravitation and the Cosmos, The Pennsylvania State University, University Park, PA 16802, USA
     	 \and
     	 	Department of Physics and Astronomy, University of Utah, 115 S 1400 E, Salt Lake City, UT 84112, USA
     	 \and
     	 	Steward Observatory, University of Arizona, 933 North Cherry Avenue, Tucson, AZ 85721
     	 \and
     	 	Faculty of Sciences, Department of Astronomy and Space Sciences, Erciyes University, 38039, Kayseri, Turkey
     	 \and
     	 	Institute of Theoretical Physics, University of Zurich, Winterthurerstrasse 190, 8057 Zurich, Switzerland
     	 \and
     	 	Department of Astronomy, University of Florida, Gainesville, FL 32611-2055, USA
     	 \and
     	 	Institute of Astronomy, University of Cambridge, Madingley Road, Cambridge CB3 0HA, UK
     	 \and
     	 	McWilliams Center for Cosmology, Department of Physics, Carnegie Mellon University, Pittsburgh, PA 15213, USA
     	 \and
     	 	Instituci\'o Catalana de Recerca i Estudis Avan\c cats, Barcelona, Catalonia
     	 \and
     	 	Institut de Ci\`encies del Cosmos, Universitat de Barcelona (UB-IEEC), Barcelona, Catalonia
     	 \and
     	 	Department of Astronomy, Ohio State University, Columbus, OH, 43210, USA
     	 \and
     	 	Institute of Cosmology and Gravitation, University of Portsmouth, Dennis Sciama building, Portsmouth P01 3FX, UK
     	 \and
     	 	Instituto de F\'\i sica, UFRGS, Caixa Postal 15051, Porto Alegre, RS - 91501-970, Brazil
     	 \and
     	 	Laborat\'orio Interinstitucional de e-Astronomia, - LIneA, Rua Gal.
Jos\'e Cristino 77, Rio de Janeiro, RJ - 20921-400,Brazil
     	 \and
     	 	INAF - Osservatorio Astronomico di Trieste, Via G. B. Tiepolo 11, I-34131 Trieste, IT
     	 \and
     	 	INFN/National Institute for Nuclear Physics, Via Valerio 2, I-34127 Trieste, IT
     	 \and
     	 	Center for Cosmology and Particle Physics, New York University, New York, NY 10003, USA
     	 \and
     	 	PITT PACC, Department of Physics and Astronomy, University of Pittsburgh, Pittsburgh, PA 15260, USA
     	 \and
     	 	Department of Astronomy and Astrophysics, University of Chicago, 5640 South Ellis Avenue, Chicago, IL 60637, USA
     	 \and
     	 	Enrico Fermi Institute, University of Chicago, 5640 South Ellis Avenue, Chicago, IL 60637, USA
       }

   \date{Received xxx; accepted xxx}

  \abstract{We present the Data Release 10 Quasar (DR10Q) catalog from the Baryon Oscillation Spectroscopic Survey
(BOSS) of the Sloan Digital Sky Survey III.
The catalog includes all BOSS objects that were targeted as quasar 
candidates  during the first 2.5 years of the survey and that are confirmed as quasars via visual inspection of the spectra,
 have luminosities $M_{\rm i}$[z=2]~$<$~$-$20.5 (in a $\Lambda$CDM cosmology with $H_0$ = 70 km~s$^{-1}$~Mpc$^{-1}$, 
$\Omega_{\rm M}$ = 0.3, and $\Omega_{\Lambda}$ = 0.7), and either display at least one emission line with a 
full width at half maximum (FWHM) larger than 500~km~s$^{-1}$
or, if not, have interesting/complex absorption features. The catalog also includes known quasars (mostly from SDSS-I and II) 
that were reobserved by BOSS.
The catalog contains 166,583 quasars (74,454 are new discoveries since SDSS-DR9)
 detected over 6,373~deg$^{2}$ with robust identification 
and redshift  measured by a combination of principal component eigenspectra.
The number of quasars with $z>2.15$ (117,668) is $\sim$5 times greater than the number of $z>2.15$ quasars known prior to BOSS.
Redshifts and FWHMs are provided for the strongest emission lines (C~{\sc iv}, C~{\sc iii}, Mg~{\sc ii}). 
The catalog  identifies 16,461 broad absorption line quasars  and gives their characteristics.
For each object, the catalog presents 
five-band (\textit{u}, \textit{g}, \textit{r}, \textit{i}, \textit{z}) CCD-based photometry with typical accuracy of 0.03 mag  
and information on the optical morphology and selection method. 
The catalog also contains X-ray, ultraviolet, near-infrared, and radio emission properties of the quasars, 
when available, from other large-area surveys. 
The calibrated digital spectra cover the wavelength region 3,600-10,500~\AA\ at 
a spectral resolution in the range 1,300~$<$~$R$~$<$~2,500; the spectra can be retrieved from the SDSS Catalog Archive Server.
We also provide a supplemental list of an additional 2,376 quasars that have been identified  among the galaxy targets of the SDSS-III/BOSS.
}

   \maketitle
 
%
\section{Introduction}
\label{s:Introduction}

Quasars have become a key ingredient in our understanding of cosmology and galaxy evolution ever since their discovery \citep{schmidt1963}.
With the advent of large and dedicated surveys such as the Sloan Digital Sky Survey \citep[SDSS;][]{york2000} and the 2dF Quasar Redshift Survey \citep[2QZ;][]{croom2001}, the number of known quasars has rapidly increased.
The first two stages of the SDSS discovered and spectroscopically confirmed more than 105,000 quasars \citep{schneider2010}, mainly at low redshift ($z \leq 2$).
The Baryonic Oscillation Spectroscopic Survey \citep[BOSS;][]{dawson2013}, the main dark time survey of the third stage of the SDSS \citep[SDSS-III;][]{eisenstein2011}, aims to measure the expansion rate of the Universe using the clustering of luminous red galaxies at $z \sim 0.7$ \citep{anderson2012} and the clustering of the intergalactic medium (IGM) at $z \sim 2.5$ \citep{busca2013,slosar2013}.
To make such a measurement, the BOSS quasar target selection \citep{ross2012} is designed to reach a sky density of $z \geq 2.15$ quasars of at least 15 ${\rm deg^{-2}}$.
The first two years of observation of SDSS-III/BOSS were released as part of the ninth data release \citep{DR9} whose associated quasar catalog \citep[DR9Q;][]{paris2012} contained 87,822 unique quasars, 61,931 having $z > 2.15$, over an area of 3,275 ${\rm deg^2}$.
\\

This paper presents the SDSS-DR10 quasar catalog, denoted DR10Q, which compiles all the spectroscopically-confirmed quasars 
identified during the first three years of BOSS operations and released as part of the SDSS tenth data release \citep{DR10}.
This catalog contains quasars targeted by the main quasar target selection \citep{ross2012}, the BOSS ancillary programs \citep{dawson2013} 
and serendipitous discoveries in the galaxy targets.
It contains 166,583 unique quasars, including 117,668 with $z > 2.15$, over an area of 6,373 ${\rm deg^2}$.
In this paper, we summarize the procedures used to compile DR10Q and describe the changes relative to DR9Q \citep{paris2012}. 

In \Sec{survey}, we summarize the target selection and observations.
We describe the construction of the DR10Q catalog in \Sec{Construction_Catalog}.
We summarize the general properties of the present sample in \Sec{DR10summary} and describe its detailed contents in \Sec{Catalog_description}. We conclude in \Sec{Summary}.
In addition, we explain the programs used to target quasars (main selection and ancillary programs) in \App{QTSflag} and 
the format of the file containing the results of the visual inspection of spectra
 used to derive the DR10Q catalog in \App{supersetDR10Q}.
This information is made available on the
SDSS public website http://www.sdss3.org/dr10/algorithms/qso\_catalog.php.
\\ 

In the following, we use a $\Lambda$CDM cosmology with ${\rm H_0 = 70 \ km \ s^{-1}, \ \Omega _M = 0.3, \ \Omega _{\Lambda} = 0.7}$ \citep{Spergel2003}.
We call a quasar an object with a luminosity ${\rm M_i\left[z=2\right] < -20.5}$ that either displays at least one emission line with 
FWHM$>$500 ${\rm km \ s^{-1}}$  or, if not, has specific absorption features that can be securely identified as quasars due to the \lya\ forest or BAL troughs.
This definition is the same as in DR9Q \citep{paris2012}. In the following, all magnitudes are PSF magnitudes \citep{stoughton2002} and are corrected for Galactic extinction using the maps from \cite{schlegel1998}.

%
%
\section{Observations}
\label{s:survey}

A surface density of $\sim$15 quasars per ${\rm deg^2}$ in the redshift range 2.15-3.5 is necessary to measure the BAO scale
in the \lya\ forest at $z \sim 2.5$ \citep{mcdonald2007}.
SDSS-III/BOSS is a five-year program that aims at observing over 160,000 $z > 2.15$ quasars over 10,000 ${\rm deg^2}$ in order to reach a precision of 4.5\% on the angular diameter distance 
and 2.6\% on the Hubble constant at $z\sim 2.5$ \citep{eisenstein2011}.
The first detection of the baryon acoustic oscillation (BAO) signal in the clustering of the IGM was obtained from the spectra of the $\sim$60,000 high redshift quasars
listed in the DR9Q catalog \citep{busca2013,slosar2013}.

	\subsection{Imaging data}

Quasar target selection for BOSS is based on imaging released in SDSS-DR8 \citep{DR8}; it  is the same as in SDSS-I/II with an extension to the South Galactic Cap (SGC).
These data were gathered using a dedicated 2.5 m wide-field telescope \citep{gunn2006}
to collect light for a camera with 30 2k$\times$2k CCDs \citep{gunn1998} over five broad bands - \textit{ugriz} \citep{fukugita1996}; 
this camera has imaged 14,555 unique square degrees of the sky, including contiguous areas of $\sim$7,500 ${\rm deg^2}$ in the North Galactic Cap (NGC) and $\sim$3,100 ${\rm deg^2}$ in the SGC that comprise the uniform ``Legacy" areas of the SDSS \citep{DR8}. The imaging data were acquired on dark photometric nights
of good seeing \citep{hogg2001}. Objects were detected and their properties were measured \citep{lupton2001,stoughton2002}
 and calibrated photometrically 
\citep{smith2002,ivezic2004,tucker2006,padmanabhan2008}, and astrometrically \citep{pier2003}.
The BOSS limiting magnitude for target selection is $r \leq 21.85$ or $g \leq 22$.

	\subsection{Target selection}
	\label{s:QTS}

The measurement of clustering of the IGM is independent of the properties of background quasars, but does depend on the surface density of quasar lines of sight. 
Hence, the target selection for the quasars used for \lya\ forest cosmology does not have to be uniform, but should maximize the surface density of high-z quasars. 
However, there is also the desire to perform demographic measurements using a uniformly-selected quasar sample. 
Thus, a [composite] strategy that mixes a heterogeneous selection to maximize the surface density of $z>2.15$  quasars, with a uniform subsample has been adopted by SDSS-III/BOSS \citep{ross2012}.

On average, 40 fibers per ${\rm deg^2}$ are allocated to the quasar project. 
Approximately half of these are selected through the target selection algorithm intended to create a uniform (``CORE") sample. 
After testing during the first year of BOSS observations, the CORE selection was chosen to be the XDQSO method \citep{bovy2011} 
The other half of the fibers are dedicated to quasar candidates that are used to maximize the surface density [at the cost of homogeneity] of high-redshift quasars: the ``BONUS" sample. 
Several different methods (a neural network combinator: \cite{yeche2010}; a Kernel Density Estimator, KDE: \cite{richards2004,richards2009}; a likelihood method: \cite{kirkpatrick2011}, 
and the XDQSO method with lower likelihood than in the CORE sample) were implement to select the BONUS quasar targets. 
Where available, near-infrared data from UKIDSS \citep{lawrence2007}, ultraviolet data from GALEX \citep{martin2005}, along with deeper coadded imaging 
in overlapping SDSS runs \citep{DR8}, were also incorporated to increase the high-z quasar yield \citep{Bovy2012}. 
Point sources that match the FIRST radio survey \citep[July 2008 version; ][]{becker1995} with $(u-g) > 0.4$ (to filter out $z < 2.15$ quasars) are 
always included in the quasar target selection.

Previously known quasars were also re-targeted\footnote{During the first two years of observations, we re-targeted $z > 2.15$ known quasars only. 
We have extended the re-observation to known $z > 1.8$ quasars since Year 3.} to take advantage of the improved throughput of the SDSS spectrographs. The sample of previously known quasars were drawn from multiple sources. These include: SDSS-DR7 quasar catalog \citep{schneider2010}; 
the 2dF QSO redshift survey \citep[2QZ;][]{croom2004}; the 2dF-SDSS LRG and QSO survey \citep[2SLAQ;][]{croom2009}; the AAT-UKIDSS-DSS (AUS) survey, 
and the MMT-BOSS pilot survey \citep[Appendix C in][]{ross2012}. 
Quasars observed at high spectral resolution by VLT/UVES and Keck/Hires were also included in the sample.

In addition to the main survey, about 5\% of the SDSS-III/BOSS fibers are allocated to ancillary programs. 
They are described in the Appendix and Tables~6 and 7 of \cite{dawson2013} and \S 4.2 of \cite{DR10}.
The full list of ancillary programs (and their target selection bits) that target quasars and thus, are included in the present catalog, 
is provided in \App{QTSflag}.

	\subsection{Spectroscopy}

\begin{figure}[htbp]
	\centering{\includegraphics[angle=-90,width=\linewidth]{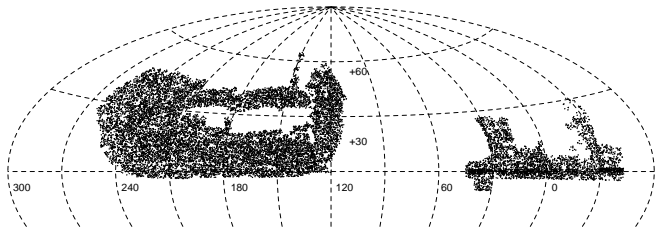}}
\caption{Distribution on the sky of the SDSS-DR10/BOSS spectroscopy in J2000 equatorial coordinates. Note that the RA~=~0 point is offset from the center. 
}
\label{fig:SkyCoverage}
\end{figure}

Quasar targets selected by the various target selection algorithms are spectroscopically observed with the BOSS spectrographs whose spectral resolution varies from  $\sim$1,300 at 3,600~\AA~ to 2,500 at 10,000~\AA~ \citep{smee2013}.
Spectroscopic observations are taken in a series of at least three 15-minute exposures.
Additional exposures are taken until the squared signal-to-noise ratio per pixel, (S/N)$^2$, reaches the survey-quality threshold for each CCD.
These thresholds are ${\rm (S/N)^2} \geq 22$ at $i$-band magnitude 21 for the red camera
and ${\rm (S/N)^2} \geq 10$ at $g$-band magnitude 22 for the blue camera (Galactic extinction-corrected magnitudes). 
The spectroscopic reduction pipeline for BOSS spectra is described  
in \cite{bolton2012}.
SDSS-III uses plates with 1,000 spectra each, which can overlap 
\citep{dawson2013}. A total of 1,515 plates were observed between December 2009 and July 2012, 
some plates were observed multiple times. In total, 166,583 unique quasars have been spectroscopically confirmed based on our visual inspection as we describe below. \Fig{SkyCoverage} shows the observed sky area.
The total area covered by the SDSS-DR10 is 6,373 ${\rm deg^2}$.

%
\section{Construction of the DR10Q catalog}
\label{s:Construction_Catalog}

The SDSS-DR10 quasar catalog is built upon the visual inspection of the spectra of all the objects targeted as quasars by SDSS-III/BOSS 
and all the objects classified robustly as $z\geq 2$ quasars by the SDSS pipeline \citep{bolton2012} among the galaxy targets.

	\subsection{Visual inspection process}

The visual inspection process 
is fully described in Section 3 of \cite{paris2012}; we briefly summarize the steps here.

The spectra of quasar candidates are reduced and the SDSS pipeline\footnote{The software used is called idlspec2d and is publicly available.
The current version is v5\_5\_12. Details can be found at
http://www.sdss3.org/dr10/software and in \cite{bolton2012}} provides a classification  
({\tt QSO}, {\tt STAR} or {\tt GALAXY}) and a redshift.
Each spectrum is fit with a library of star templates, a PCA decomposition of galaxy spectra and a PCA decomposition of quasar spectra.
Each class of templates is fit over a range of redshifts: galaxies from $z = -0.01$ to $1.00$, quasars from $z = 0.0033$ to $7.00$, and stars from $z = -0.004$ to $0.004$ ($\pm  1,200 \ {\rm km \ s^{-1}}$).
The combination of redshift and template with the overall best fit (in terms of the lowest reduced chi-squared) is adopted as the pipeline classification and redshift measurement.
The redshift is accompanied by a flag ({\tt ZWARNING}); when it is zero, the pipeline has high confidence in its estimate.  
We use this output as a first estimate for the visual inspection.

Through a dedicated website, quasar candidates are separated between low-z quasars ($z<2$), high-z quasars ($z\geq 2$), stars and ``others" based on the initial SDSS pipeline classification.
The spectra of the objects in each list are then visually inspected and each object is classified as {\tt QSO}, {\tt Star}, {\tt Galaxy} if the identification and the redshift are certain, {\tt QSO\_Z?} if the object is clearly a quasar but the redshift is uncertain, {\tt QSO\_?} and {\tt Star\_?} if we have clues to the correct classification but are not certain, {\tt ?} if we cannot certify the classification of an object at all and {\tt Bad} when the signal-to-noise ratio is too low to identify the objects.
The distinction between {\tt ?} and {\tt Bad} becomes quite subjective as the signal-to-noise ratio decreases.

In addition to the quasar candidates, we also visually inspected the spectra of some objects among the galaxy targets. 
While confirming the identification of all the quasar targets is important to understand and improve the various quasar target selection algorithms, objects are inspected among the galaxy targets to get back the maximum number of serendipitous quasars in the full SDSS-III/BOSS spectroscopic sample.
Moreover, these objects usually have low signal-to-noise ratio, and thus, their identification is  more difficult.
For these reasons, the visual inspection strategy is slightly different for this class of objects, and we adopted a binary classification: either an object is a {\tt QSO} or this is \textit{not a quasar}.

	\subsection{Definition of the DR10Q parent sample}

\begin{table*}
\centering                          
\begin{tabular}{l r r r}        
\hline                
\hline
 Classification & \# pipeline & \# pipeline           & \# visual \\   
                &             & with {\tt ZWARNING}=0 &  inspection          \\
\hline 
\hline                       
{\tt QSO}                           & 188,705     & 162,857                & 166,583   \\
{\tt QSO} with $z>2.15$             & (124,062)   & (115,429)              & (117,668) \\
{\tt QSO\_?}                        & -           & -                      & 1,370    \\
{\tt QSO\_Z?}                       & -           & -                      & 545      \\
{\tt Galaxy}                        & 20,627      & 14,512                 & 10,267    \\
{\tt Star}                          & 112,397     & 83,429                 & 133,557   \\
{\tt Star\_?}                       & -           & -                      & 1,519      \\
{\tt Bad}                           & -           & -                      & 3,257    \\
{\tt ?}                             & -           & -                      & 1,082       \\
Missing$^a$                         & -           & -                      & 150        \\
Not a quasar (galaxy targets)       & -           & -                      & 3,398    \\
\hline
Total                               & 321,729     & 260,798                & 321,729  \\
\hline       
\multicolumn{4}{l}{$^a$ Quasars not visually inspected because of bad photometric information}                         
\end{tabular}
\caption{
	Number of objects identified as such by the pipeline with any  {\tt ZWARNING} value  (second column) and 
with {\tt ZWARNING}~=~0 (third column), and after the visual inspection (fourth column).
	}           
\label{t:DR10nb}      
\end{table*}

We selected quasar candidates among the objects targeted by the main quasar target selection \citep[labelled as {\tt BOSS\_TARGET1} in the SDSS database, ][]{ross2012}, the ancillary programs that target quasars \citep[{\tt ANCILLARY\_TARGET1} and {\tt ANCILLARY\_TARGET2}, see the Appendix of ][]{dawson2013} and objects identified as {\tt QSO} with {\tt z}$>$2 and {\tt ZWARNING}=0 by the SDSS pipeline or identified as {\tt GALAXY} with the subclass {\tt BROADLINE}.
This leads to a superset of 316,947 quasar targets (i.e. objects targeted for spectroscopy as quasars) and 4,782 possible serendipitous quasars from the galaxy targets for a total number of objects of 321,729.
The spectra of 321,579 of these have been visually inspected, and 150 quasar targets are missing because of bad photometric information.
166,583 objects have been identified as {\tt QSO}, 545 as {\tt QSO\_Z?}, 1,370 as {\tt QSO\_?}, 10,267 as {\tt Galaxy}, 133,557 as {\tt Star}, 1,519 as {\tt Star\_?}, 1,082 as {\tt ?} and 3,257 as {\tt Bad}.
The result of the visual inspection is given in \Tab{DR10nb}.

This quasar catalog lists all the firmly confirmed quasars ({\tt QSO} and {\tt QSO\_BAL}, i.e. {\tt QSO} showing broad absorption lines, only).
About 10\% of these quasars have been observed spectroscopically several times \citep{dawson2013}.
These repeat observations are often useful to confirm the nature of objects with
low S/N spectra. However, we did not attempt to co-add these data
mostly because they are often of quite different S/N. 

Together with the quasar catalog, we provide the full result of the visual inspection (e.g. the identification of all visually scanned objects)
with the whole superset from which we derived this catalog.
We encode the identification using two parameters {\tt z\_conf\_person} and {\tt class\_person}. The relation between these two parameters 
and the output from the visual inspection is given in \Tab{VI_PIPE}. 
We refer the reader to \App{supersetDR10Q} to find the detailed format of the fits file that contains the DR10Q superset.

Finally, in the course of various tests for the inclusion of galaxy targets in the visual inspection process, we gathered a sample of serendipitous quasars.
The parent sample of these objects is not well defined as the visual inspection strategy for the galaxy targets has changed several times.
Hence, we provide them in the form of a supplementary list that contains 2,376 quasars.

%
\begin{table}
\centering                        
\begin{tabular}{c c c c c}        
\hline\hline
                & \multicolumn{4}{c}{z\_conf\_person}  \\                                        
                & 0                  & 1                  & 2             & 3        \\
class\_person   &                    &                    &               &          \\
\hline
0               & Not inspected      & {\tt ?}            & -             & -        \\
1               & -                  & -                  & {\tt Star\_?} & {\tt Star}     \\
2               & Not a quasar $^a$  & -                  & -             & -              \\
3               & -                  & {\tt QSO\_?}       & {\tt QSO\_Z?} & {\tt QSO}      \\
4               & -                  & -                  & -             & {\tt Galaxy}   \\
30              & -                  & -                  & -             & {\tt QSO\_BAL} \\
\hline   
\multicolumn{5}{l}{$^a$ Galaxy targets}                         
\end{tabular}
\caption{The visual inspection classification, corresponding to the combination of the {\tt class\_person} (first column) and {\tt z\_conf\_person} (first row) values provided in the superset file described in \App{supersetDR10Q}. 
}
\label{t:VI_PIPE}
\end{table}

%
\section{Summary of sample}
\label{s:DR10summary}
\subsection{Broad view}

\begin{figure*}[htbp]
	\centering{
		\begin{minipage}{.45\linewidth}
			\centering{\includegraphics[width=\linewidth]{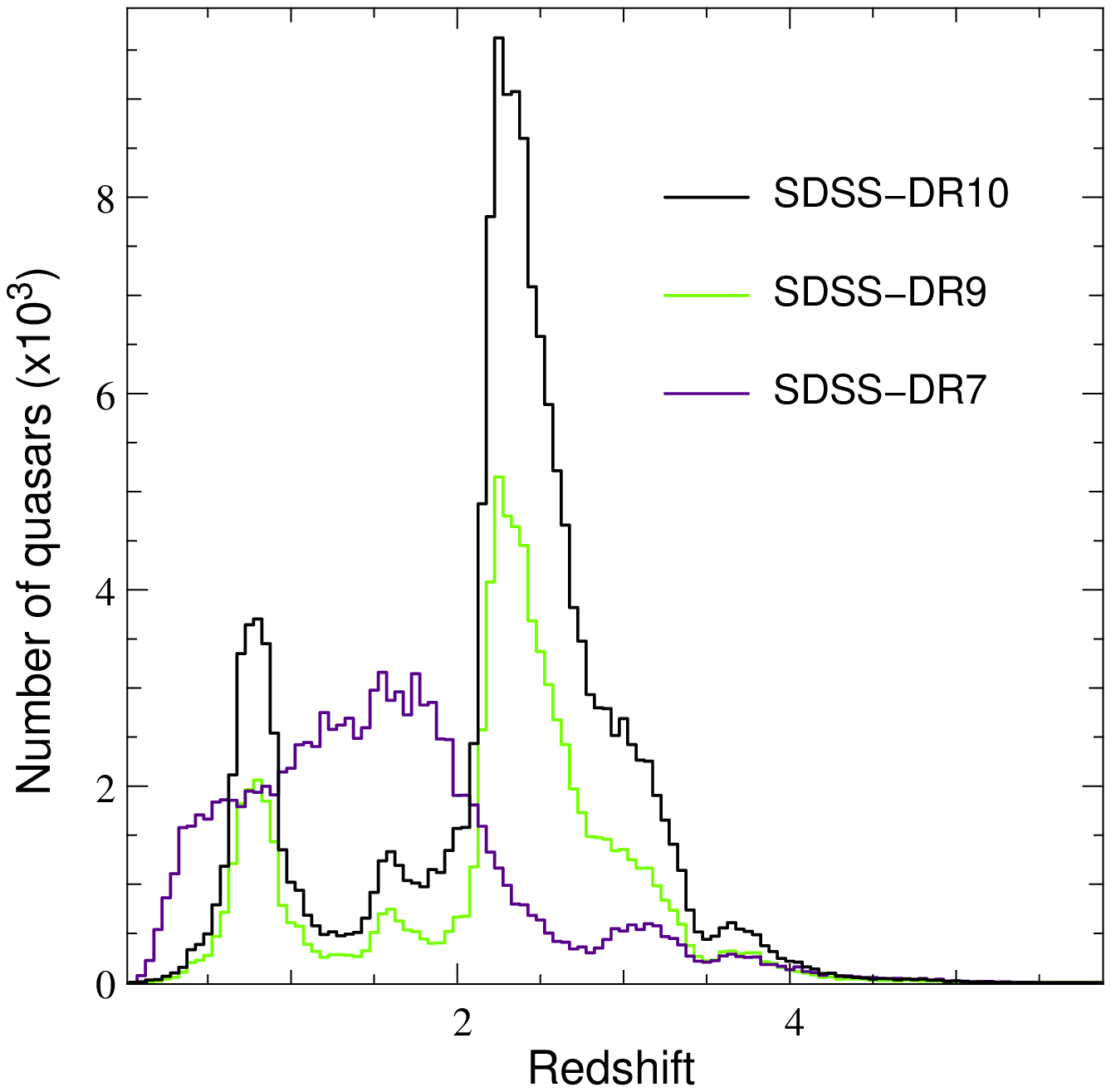}}
		\end{minipage}
		\hfill
		\begin{minipage}{.45\linewidth}
			\centering{\includegraphics[width=\linewidth]{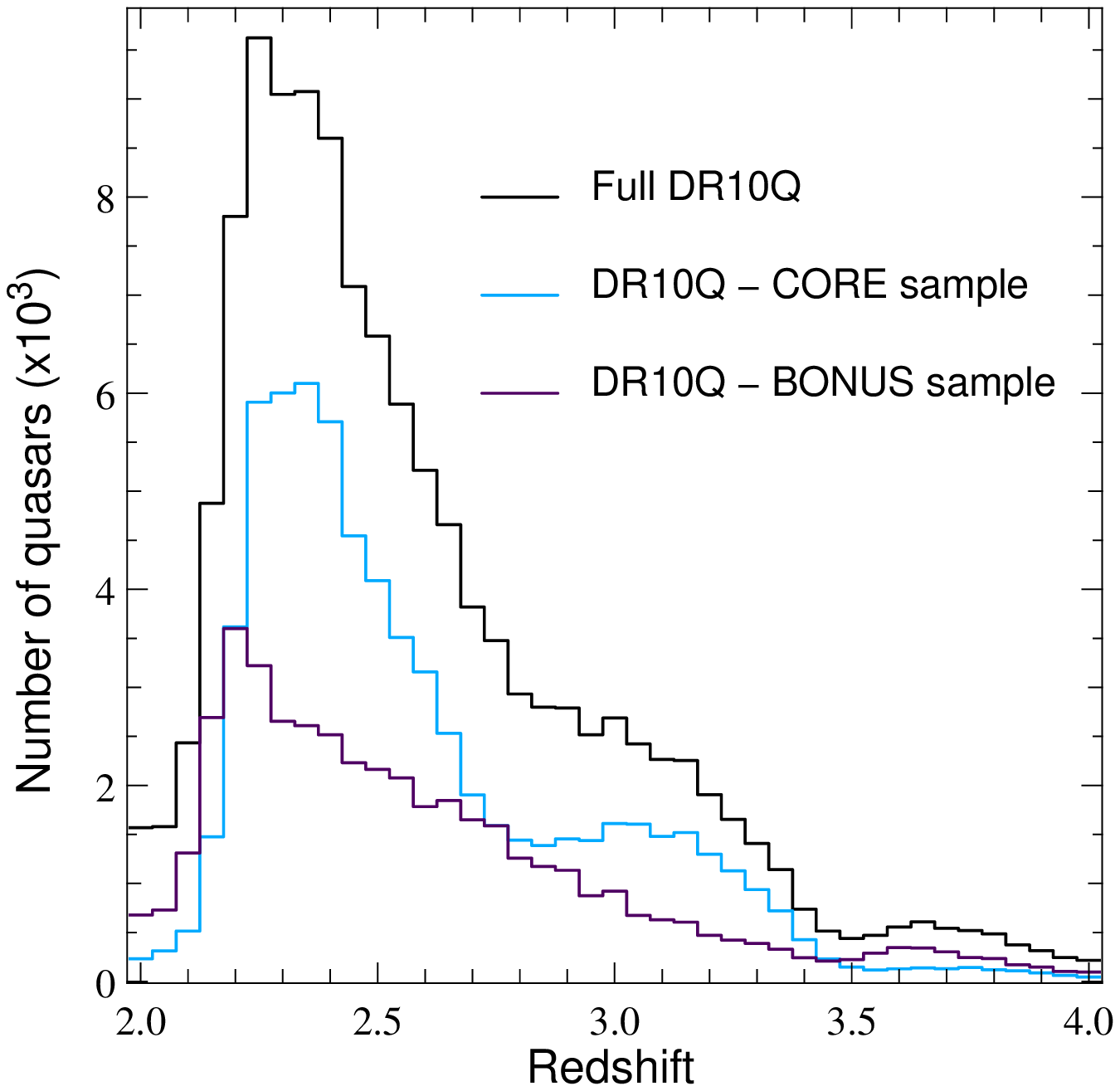}}
		\end{minipage}
		}
	\caption{
	\textit{Left panel: }	
	Redshift distribution of the SDSS-DR10 (black histogram), SDSS-DR9 \citep[green histogram; ][]{paris2012} and SDSS-DR7 \citep[purple histogram; ][]{schneider2010} quasars over the redshift range 0-6.
	\textit{Right panel: }
	Redshift distribution of the full SDSS-DR10 quasar sample (black histogram), and the SDSS-DR10 CORE (blue histogram) 
and BONUS (purple histogram) samples over the redshift range 2 to 4.
Note that the full DR10Q sample includes CORE, BONUS and quasars observed as part of ancillary programs.
	}
	\label{fig:zdistriDR10}
\end{figure*}

\begin{figure*}[htbp]
	\centering{\includegraphics[angle=-90,width=.9\linewidth]{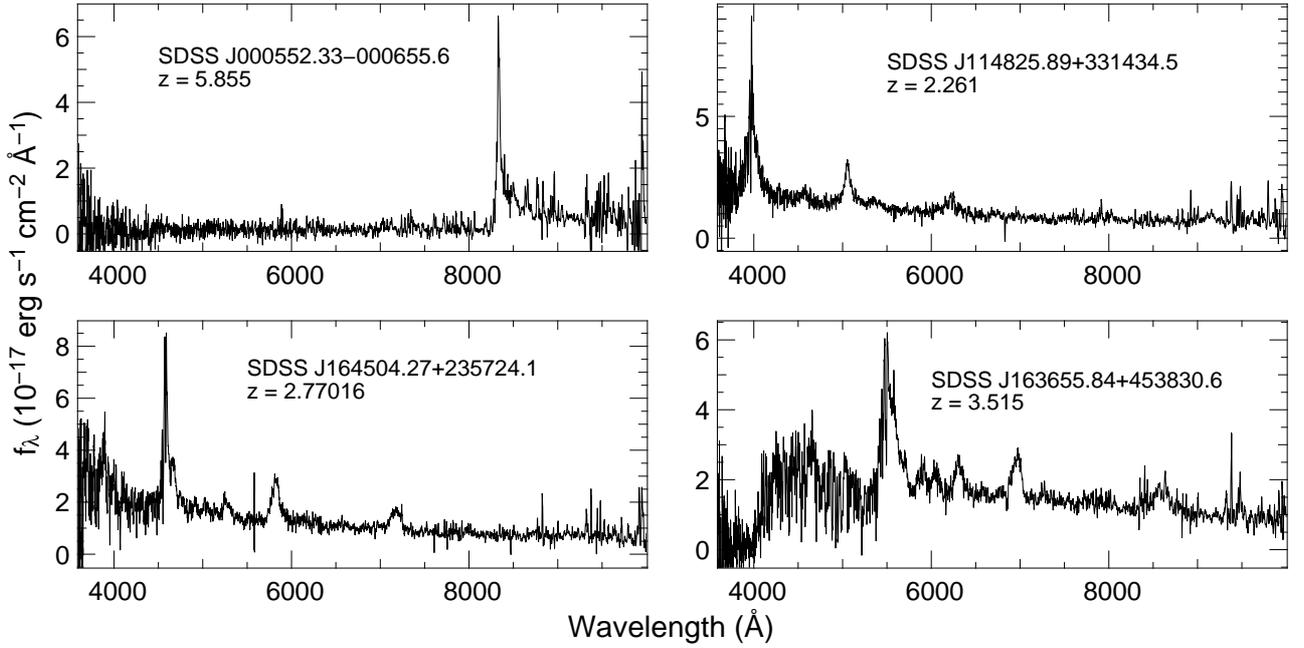}}
\label{fig:exspec}
\caption{The highest redshift quasar of the SDSS-DR10 quasar catalog is shown in the upper left panel. Note that this quasar was already in DR9Q. The three other panels show typical examples of quasar spectra at different redshift.}
\end{figure*}

\begin{figure}[htbp]
	\centering{\includegraphics[width=.9\linewidth]{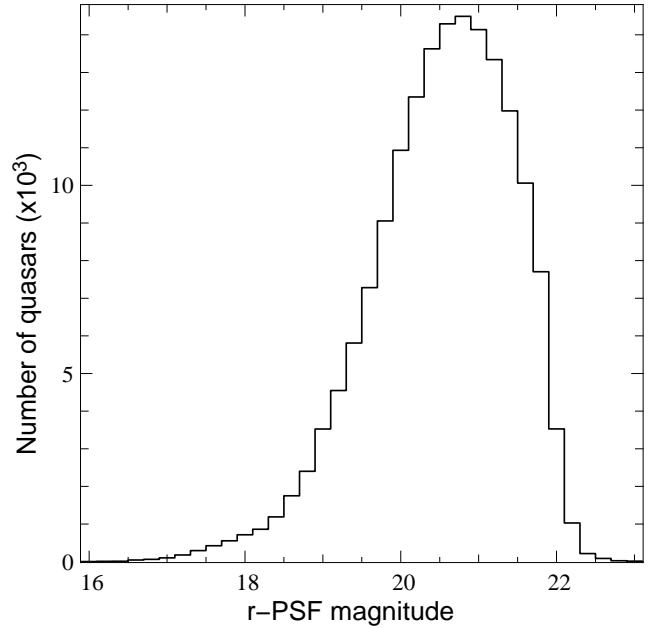}}
	\caption{
	Distribution of $r$-PSF magnitude (corrected for Galactic extinction) of the quasars included in the DR10Q catalog.	
	}
	\label{fig:rmagDistri}
\end{figure}

\begin{figure}[htbp]
	\centering{\includegraphics[width=.9\linewidth]{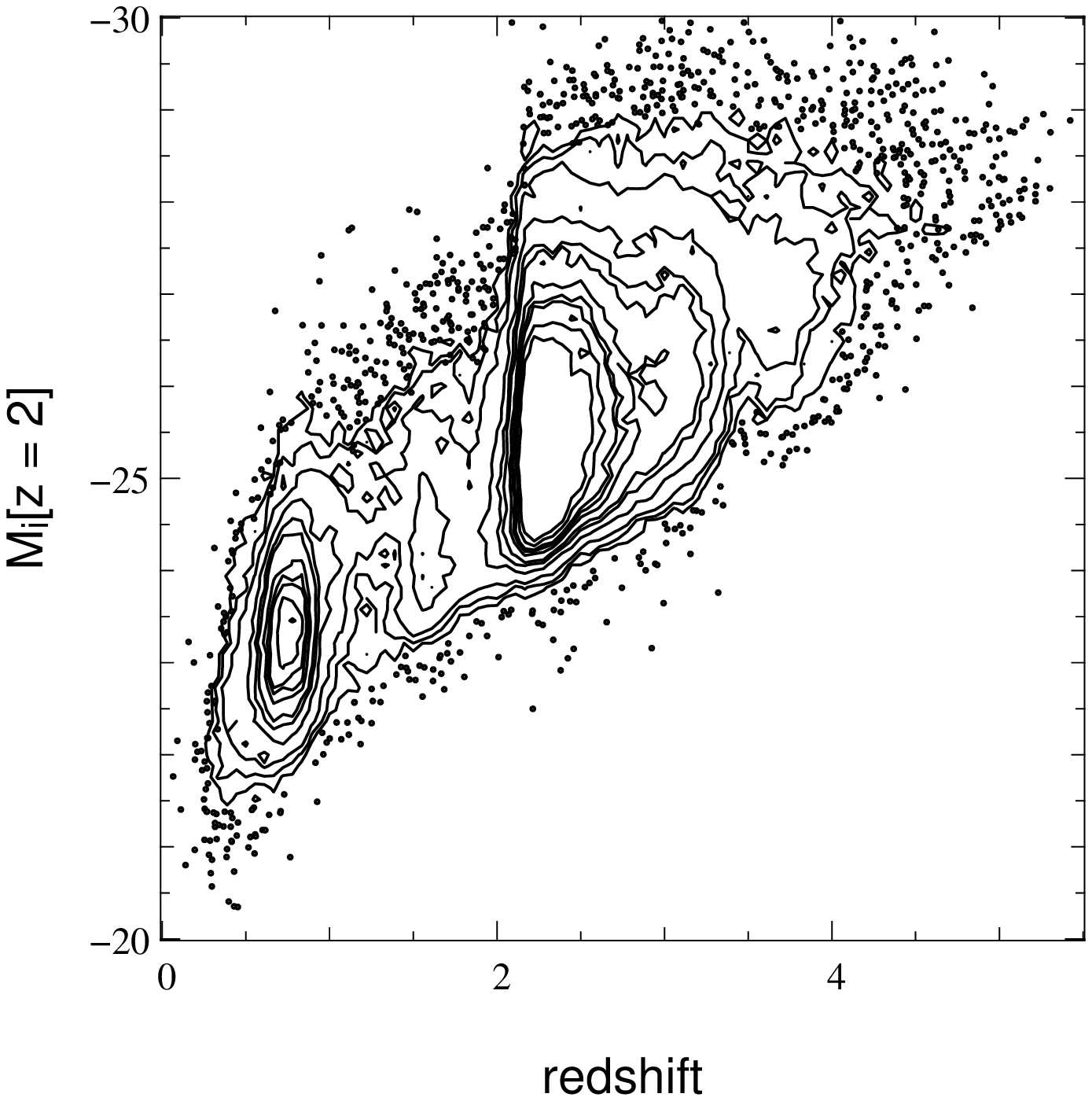}}
	\caption{
	Luminosity-redshift plane for DR10Q quasars (black contours and points).
	The luminosity assumes ${\rm H_0} = 70 \ {\rm km \ s^{-1} \ Mpc^{-1}}$ and the k-correction is given by \cite{richards2006} 
who consider $K(z=2) = 0$. Contours are drawn at constant point density.
	}
	\label{fig:LZ}
\end{figure}

The DR10Q catalog contains 166,583 unique quasars, including 117,668 with $z > 2.15$.
This sample contains about twice as many quasars as the previous release (DR9Q).
The first 2.5 years of operations of SDSS-III/BOSS cover an area of 6,373 ${\rm deg^2}$,  leading to a mean density of 16.3 $z > 2.15$ quasars per square degree.

The quasars from this catalog have redshifts between 0.053 and 5.855.
Their redshift distribution is shown in \Fig{zdistriDR10} (left panel). 
The peaks at $z \sim 0.8$ and $z \sim 1.6$ are due to known degeneracies in the SDSS color space.
The SDSS-DR9 (green histogram) and SDSS-DR10 (black histogram) redshift distributions are scaled versions of one another. 
The right panel of \Fig{zdistriDR10} displays the same distribution but in the redshift range 
of interest for BOSS ($2.00 \leq z \leq 4.00$) together with the distributions from 
the CORE (uniformly selected sample; blue histogram) and the BONUS (purple histogram) samples.
Typical spectra of SDSS-DR10 quasars are shown in Fig. 3.
The broad statistical properties of this sample have not changed since DR9Q. We list below the differences between the two catalogs, and refer the reader to \cite{paris2012} for a detailed description of the unchanged content.


\subsection{Differences between SDSS-DR9 and SDSS-DR10 quasar catalogs}

	\subsubsection{Quasars that were in DR9Q but not in DR10Q and quasars that could have been in DR9Q but were not}

The DR9Q catalog \citep{paris2012} contained 87,822 unique quasars of which 86,952 are also part of the present catalog.
Two plates (3698 and 5369) were accidentally included in DR9 \citep{DR9}, although they did not fulfill the minimum S/N
conditions \citep[see ][]{dawson2013}. They have been downgraded to ``bad plates'', and therefore not included in DR10 \citep{DR10}. 
These plates have been re-drilled, and re-observed after the DR10 cut-off.
Hence, 90 DR9 quasars have dropped out of the present catalog.
The remaining 786 quasars that were part of DR9Q but not DR10Q were re-classified during various tests of the visual inspection process and are now part of the supplementary list as explained in \Sec{Construction_Catalog}.

In addition, the identification of 56 objects have changed between DR9Q and DR10Q. This modification is due to systematic checks performed in the past year on 
objects classified  as {\tt QSO\_?} and {\tt ?}.  These checks are necessary especially since objects can be re-observed, often with 
better S/N and the data reduction pipeline constantly improves. We perform this re-inspection after the release of a new data reduction. A few mistakes have also been corrected, after feedback from users of the catalog.

	\subsubsection{Visual inspection redshift}

\begin{figure}[htbp]
	\centering{\includegraphics[width=80mm]{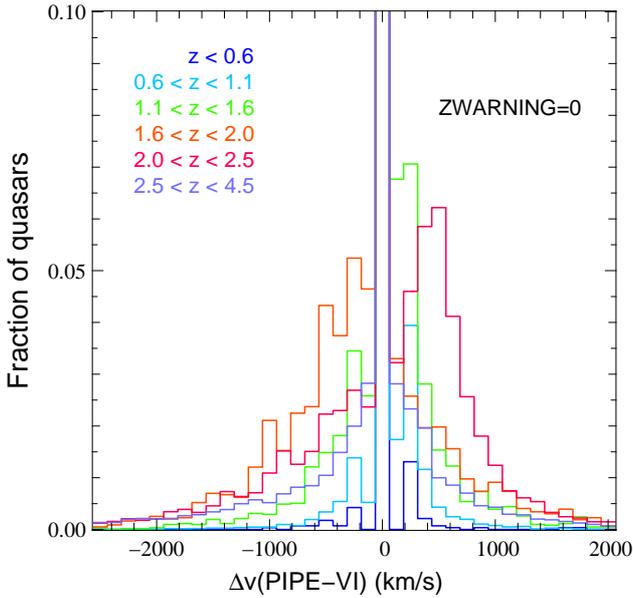}}
\caption{Distribution of velocity differences between the pipeline and the visual inspection redshift estimates in different bins of redshift. Depending on the redshift bin, the fraction of redshift that we correct increases up to $\sim$50\%.}
\label{fig:Comparison_VI}
\end{figure}

As part of the visual inspection, we use the SDSS pipeline redshift estimate ({\tt Z\_PIPE}) as a first guess, and we either confirm or correct for it when it is necessary. 
 We use the maximum of the \ion{Mg}{ii} emission line, when it is available in the spectrum, to set the visual inspection redshift ({\tt Z\_VI}). \Fig{Comparison_VI} shows the distribution of velocity differences between the pipeline and the visual inspection redshift estimates.

In the course of the checks mentioned above, some redshifts have been adjusted. In addition,
in the DR9Q catalog \citep{paris2012}, the visual inspection redshifts have a 3 digit precision because it is
not possible to reach higher precision by eye. The pipeline redshift estimate has a 5 digit precision.
In DR9Q, the truncation of the visual redshift was performed even when the visual inspection simply confirms the 
pipeline redshift. 
This approach created a systematic and arbitrary mean blueshift of 50 ${\rm km \ s^{-1}}$ 
of the visual redshift estimate relative to the pipeline estimate. In DR10Q we do not apply this truncation.

	\subsubsection{Emission Line fitting}

We fit several emission lines (\CIV , the \CIII\ complex and \MgII ) using a set of principal components. 
This fit provides a measurement of the line redshift, FWHM, blue and red HWHMs and their equivalent width.
We apply the exact same procedure as in DR9Q \citep[see Section 4 of ][]{paris2012}.

However, the set of principal components used in DR9Q was unable to reproduce emission lines with FWHM smaller than 
2,000 ${\rm km \ s^{-1}}$ \citep{alexandroff2013}, creating an arbitrary truncation of FWHM at this value.
We corrected for this defect by incorporating quasars with \CIV\ emission lines  of FWHM smaller than 
2,000 ${\rm km \ s^{-1}}$ in the sample used to derive the set of principal components.
This modification does not affect most of the measured emission line properties with a median difference for the measured \CIV\ FWHM of $\sim 50 \ {\rm km \ s^{-1}}$ between DR9Q and DR10Q.
This change does not affect the quality of the PCA redshift estimate neither. Over the $0.6 \leq z \leq 2.3$ sample, where the \ion{Mg}{ii} emission line is available in the SDSS-III/BOSS spectra, the median shift between the global PCA redshift estimate ({\tt Z\_PCA}) and the \ion{Mg}{ii} redshift is less than 5 ${\rm km \ s^{-1}}$.

\subsection{Uniform sample}
\label{s:uniform}

As in DR9Q, we provide a {\tt uniform} flag (row \#26 of Table~4) in order to identify a homogeneously selected sample of quasars.
Although the CORE sample has been designed to have a well understood, uniform, and reproducible selection function, its definition has varied over the first year of the survey. 
Thus, the nominal CORE quasars are not a uniformly-selected sample.
Areas within which all the algorithms used in the quasar target selection are uniform are called ``Chunks''.
The definition of each of them can be found in \cite{ross2012}.
After Chunk 12, CORE targets were selected with the XDQSO technique \textit{only} \citep{bovy2011}. 
These objects have {\tt uniform}~=~1.
Quasars in our catalog with {\tt uniform}~=~2 are objects that would have been selected by XDQSO if it had been the CORE algorithm prior to Chunk 12.
Objects with {\tt uniform}~=~2 are reasonably complete to what XDQSO would have selected, even prior to Chunk 12. But, as DR10Q only contains information about spectroscopically confirmed quasars, not about all {\em targets}, care must be taken to create a statistical sample from the uniform flag. See, e.g, the discussion regarding the creation of a statistical sample for clustering measurements in \cite{white2012}.
Quasars with {\tt uniform}~=~0 are not homogeneously selected CORE targets.

\subsection{Broad Absorption Line quasars}

Broad absorption lines, BALs, in quasar spectra are both visually and automatically identified and characterized as described in Section 5 of \cite{paris2012}.

In total we identified 16,461 BAL quasars during the visual inspection of the DR10 quasar candidates.
The visual inspection does not measure the width of troughs, but \CIV\ troughs are also automatically identified and characterized in the spectra of $z \geq 1.57$ quasars.
This redshift is chosen so that the whole region between \SiIV\ and \CIV\ emission lines is covered by the BOSS spectrograph.

We provide the balnicity index \citep[BI;][]{weymann1991}
and the absorption index \citep[AI;][]{hall2002} for these quasars. 
A total of 9,623 quasars with {\tt Z\_VI}$\geq$1.57 have a positive BI value and 26,232 have ${\rm AI} > 0\  {\rm km \ s^{-1}}$. 
The systematic search is performed for \CIV\ troughs \textit{only}: some quasars flagged BAL by the visual inspection may have BI~=~0 ${\rm km \ s^{-1}}$, either because the condition that troughs must be at least 2,000 ${\rm km \ s^{-1}}$ wide is not fullfilled, or because the object was flagged because of absorption lines of some other ion (e.g. \MgII , \SiIV ).
The distribution of the two indices is shown in Fig. 7.

\begin{figure*}[htbp]
	\begin{minipage}{0.45\linewidth}
		\centering{\includegraphics[width=\linewidth]{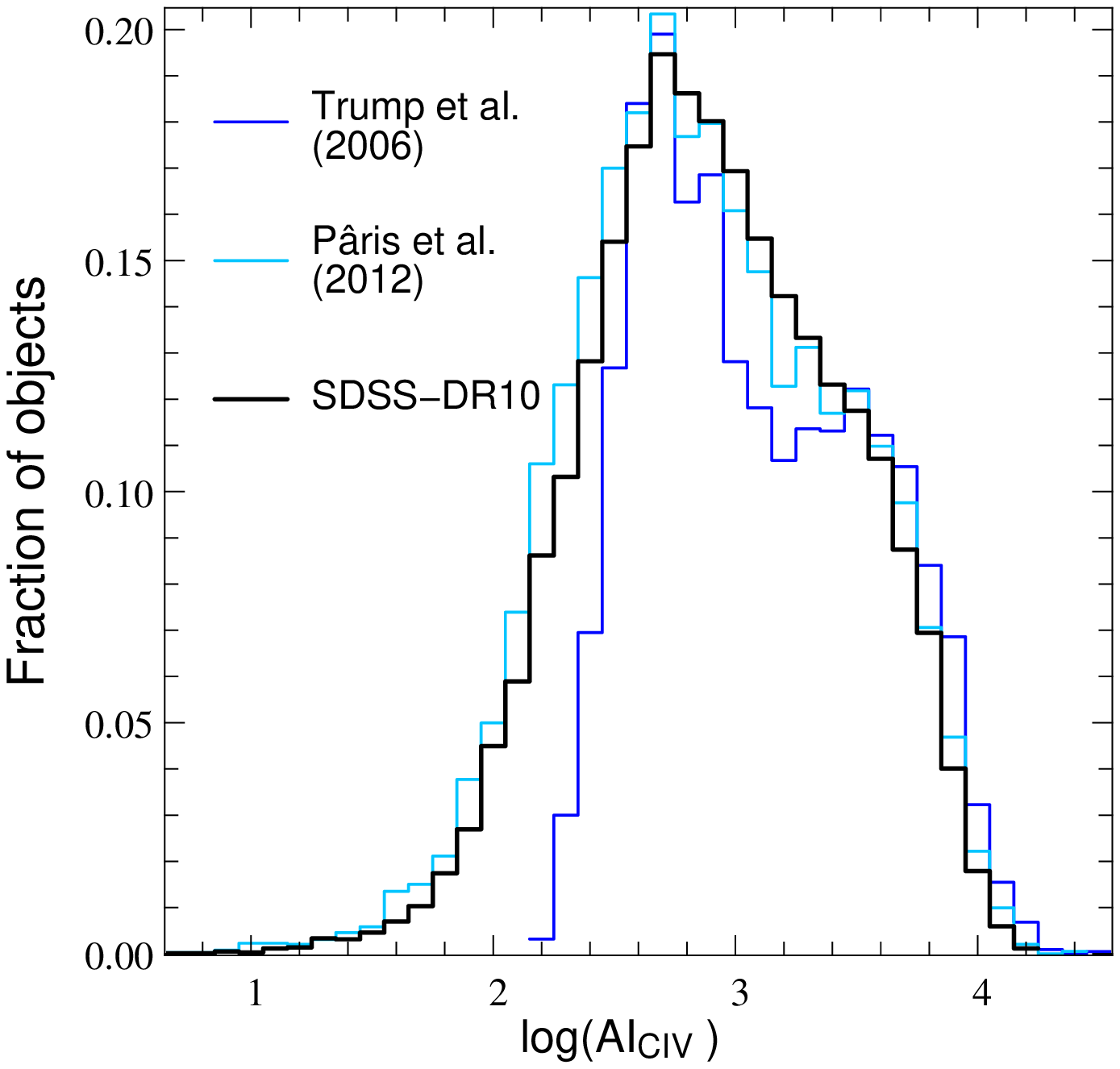}}
	\end{minipage}
	\hfill
	\begin{minipage}{0.45\linewidth}
		\centering{\includegraphics[width=\linewidth]{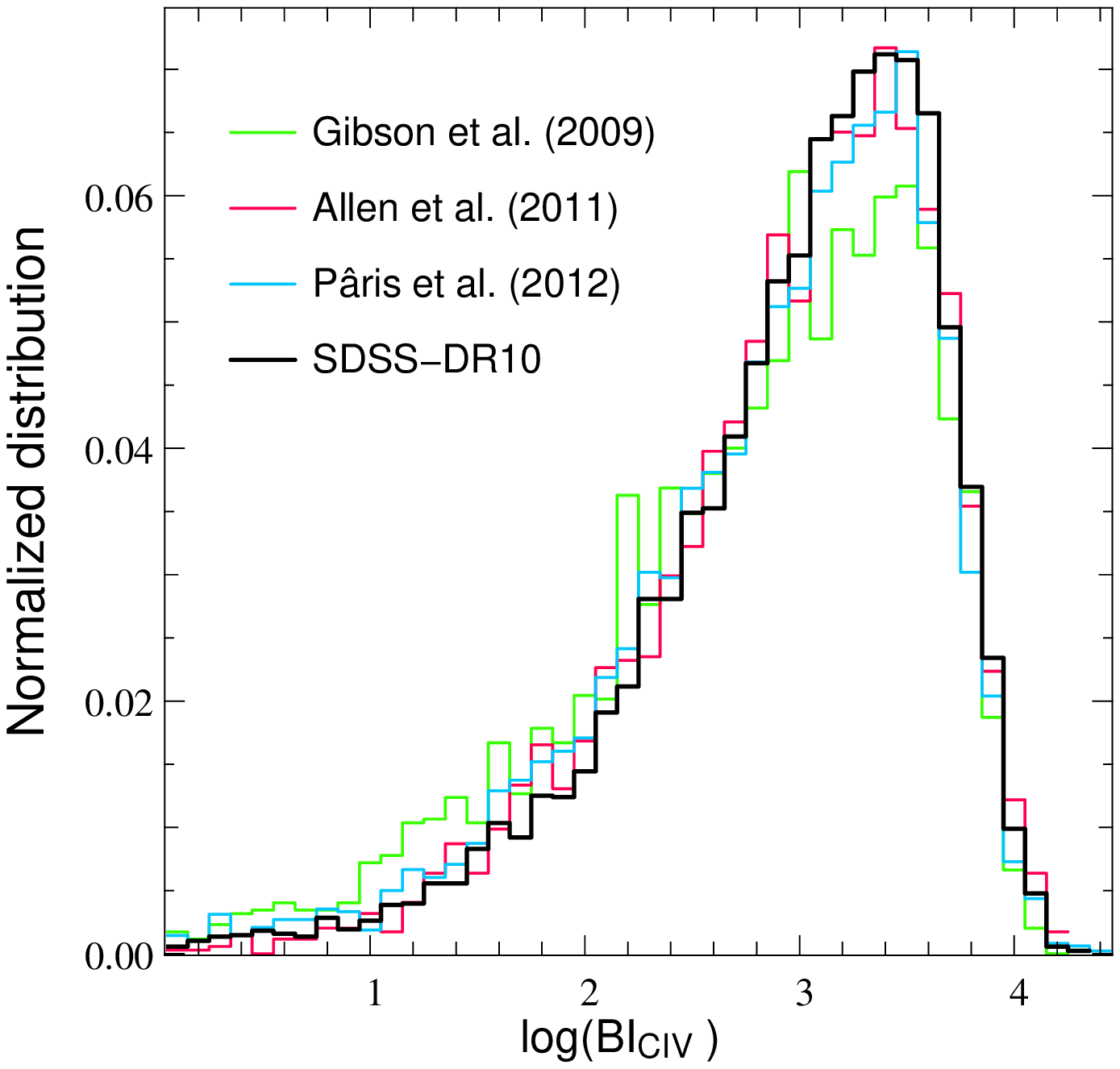}}
	\end{minipage}
\caption{\textit{Left panel:} Distribution of AI measured for \CIV\ absorption troughs from \cite{trump2006} (blue histogram), SDSS-DR9 \citep[cyan histogram; ][]{paris2012} and the present catalog (thick black histogram). The distributions are normalized for $\log {\rm AI} > 3$. \textit{Right panel:} Distribution of BI measured for the \CIV\ absorption troughs from  \cite{gibson09} (green histogram), \cite{allen2011} (red histogram), SDSS-DR9 \citep[cyan histogram; ][]{paris2012}, and the present catalog (thick black histogram).}
\end{figure*}

\subsection{Multiwavelength matching}
%

We cross-correlate the DR10Q catalog with several other surveys: the FIRST radio survey \citep{becker1995}, the Galaxy Evolution Explorer 
\citep[GALEX,][]{martin2005} survey in the UV, the Two Micron All Sky Survey \citep[2MASS,][]{cutri2003,skrutskie2006}, 
the Wide-Field Infrared Survey \citep[WISE,][]{wright2010} and the ROSAT All-Sky Survey \citep[RASS;][]{voges1999,voges2000}.
In addition, we now provide a match to the UKIRT Infrared Deep Sky Survey \citep[UKIDSS;][]{lawrence2007} and the second XMM-Newton 
serendipitous source catalog \citep{watson2009}.

	\subsubsection{FIRST}
	\label{s:FIRST}

We match the DR10Q catalog to the FIRST radio survey \citep{becker1995} using a 2\arcsec matching radius as we did in DR9Q.
We use the FIRST catalog released in February 2012, which contains an extension in the SGC which is not included in the July 2008 catalog that was used to match to DR9Q quasars.
We report the flux peak density at 20 cm and the signal-to-noise ratio of the detection.
If a quasar lies outside of the FIRST footprint, the reported flux density peak is set to $-1$.

The SDSS-III/BOSS quasar target selection \citep{ross2012} automatically includes the FIRST sources from the {\em previous} (July 2008) version of the FIRST catalog  with $(u-g) > 0.4$. 
This additional color cut is set to avoid contamination by low-redshift quasars.

There are 6,682 matches to the FIRST catalog.
	
	\subsubsection{The Galaxy Evolution Explorer (GALEX)}
\label{s:GALEX}
	
As in DR9Q, GALEX \citep{martin2005} images are force photometered (from GALEX Data Release 5) at the SDSS-DR8 centroids \citep{DR8}, such that low S/N point-spread function fluxes of objects not detected by GALEX is obtained,  
for both the FUV (1,350-1,750 \AA ) and NUV (1,750-2,750 \AA ) bands when available.

A total of 107,835 quasars are detected in the NUV band, 86,753 in the FUV band and 69,164 have non-zero fluxes in both bands.

	\subsubsection{The Two Micron All Sky Survey (2MASS)}
\label{s:2MASS}

We cross-correlate the DR10Q catalog with the All-Sky Data Release Point Source catalog \citep{skrutskie2006} using a matching radius of 2\arcsec.
We report the Vega-based magnitudes in the J, H and K-bands and their error together with the signal-to-noise ratio of the detections.
We also provide the value of the 2MASS flag rd\_flg[1], which gives the meaning of the peculiar values of the magnitude and its error for each band\footnote{see http://www.ipac.caltech.edu/2mass/releases/allsky/doc/explsup.html for more details}.

There are 2,824 matches in the catalog.

	\subsubsection{The Wide-Field Infrared Survey (WISE)}
\label{s:WISE}

As we did for DR9Q, we match the BOSS DR10Q to the 
WISE All-Sky Source Catalog\footnote{http://irsa.ipac.caltech.edu/Missions/wise.html} \citep{wright2010}. 
Our procedure is the same as in DR9Q, with a matching radius of
2.0\arcsec providing a very low-level of false positive matches \citep[see e.g. Fig. 4 of][]{krawczyk2013}. 
We find 87,849 matches from the WISE All-Sky Source Catalog. 

Here, we report the magnitudes, their associated errors, the S/N of the detection and reduced $\chi ^2$ of the profile-fitting in 
the four WISE bands centered at wavelengths of 3.4, 4.6, 12 and 22~$\mu$m. These magnitudes 
are in the Vega system, and are measured with profile-fitting photometry\footnote{see e.g. http://wise2.ipac.caltech.edu/docs/release/allsky/expsup/ sec2\_2a.html
and http://wise2.ipac.caltech.edu/docs/release/allsky/expsup/ sec4\_4c.html\#wpro}.         

However, we {\it strongly caution} the reader not to accept our
catalog matches in blind faith. We report the WISE
catalog Contamination and confusion flag, {\tt cc\_{flags}}, and as
suggested on the WISE ``Cautionary Notes" page
\footnote{http://wise2.ipac.caltech.edu/docs/release/allsky/expsup/sec1\_4b.html \#unreliab}
recommend using only thos matches with {\tt cc\_flags} = '0000' to exclude objects
which are flagged as spurious detections of image artifacts in any
band. However, our checks have found that even this criterion sometimes selects catalog objects that are obviously spurious in the
WISE imaging.  

	\subsubsection{UKIDSS}
	\label{s:UKIDSS}
	
The UKIRT Infrared Deep Sky Survey \citep[UKIDSS; ][]{lawrence2007} has observed 7,500 ${\rm deg ^2}$ in the near infrared.
UKIDSS images are force photometered at the SDSS-DR8 centroids \citep{DR8}.
UKIDSS observations acquired before March 2011 were used;
 they have been released as UKIDSS DR1-DR9.

We provide the fluxes and their associated errors, expressed in ${\rm W \ m^{-2} \ Hz^{-1}}$, in the Y, J, H and  K bands.
The conversion to the Vega magnitudes, used in 2MASS, is given by the formula:
\begin{equation}
	{\rm mag _{X}} = -2.5 \times \log \frac{f_{\rm X}}{f_{\rm 0, X} \times 10^{-26}},
\end{equation}
where X denotes the filter and the zero-point values $f_{\rm 0, X}$ are 2026, 1530, 1019 and 631 for the Y, J, H and K bands respectively.

A total of 66,649 quasars are detected in one of the four bands Y, J, H or K.
65,997 objects are detected in the Y band, 65,732 in the J band, 65,730 in H band, 66,130 in the K band and 64,181 objects have 
non-zero fluxes in the four bands.
The UKIDSS limiting magnitude is ${\rm K} \sim 18$ (for the Large Area Survey) while the 2MASS limiting magnitude in the same band is $\sim 15.3$.
This difference in depth between the two surveys explains the very different numbers of matches with DR10Q.

	\subsubsection{ROSAT all sky survey}
\label{s:ROSAT}

As in DR9Q, we cross-correlate DR10Q with the ROSAT all sky survey Faint \citep{voges2000} and Bright \citep{voges1999} source 
catalogues with a matching radius of 30".
Only the most reliable detections are included in our catalog: when the quality detection is flagged as potentially problematic, we do not include 
the match.

There are 57 matches with the Bright Source Catalog and 586 with the Faint Source Catalog.
	
	\subsubsection{XMM-Newton}
\label{s:XMM}

We cross-correlate the DR10Q catalog with the second XMM-Newton Serendipitous Source Catalog 
\citep[Third Data Release, 2XMMi-DR3;][]{watson2009} using a standard 5" matching radius.
The 2XMMi-DR3 catalog has observed 512 ${\rm deg^2}$, with possible multiple observations of a given sky area.

We report the soft (0.5--2 keV) and hard (2--10 keV) fluxes, expressed in ${\rm erg \ cm^{-2} \ s^{-1}}$ and the 
corresponding X-ray luminosities, expressed in ${\rm erg \ s^{-1}}$.
The latter are computed from the flux in each band using the visual inspection redshift ({\tt Z\_VI}) and $H_0 = 70 {\rm km \ s^{-1} \ Mpc^{-1}}, \ \Omega _M = 0.3, \ \Omega _{\Lambda} = 0.7$.

A total of 506 of the sources have been observed more than once. In these cases, the observation with the longest exposure time has been used to compute the fluxes.

The XMM-Newton telescope has a set of three X-ray CCD cameras (MOS1, MOS2 and PN).
The European Photon Imaging Camera (EPIC-PN) is preferred because of its higher quantum efficiency.
If one source has no PN observation or insufficient amount of counts, 
the average of MOS1 and MOS2 camera fluxes was used to improve statistics. 
Reported fluxes for sources with less than 30 counts in each camera are given as the
mean of the fluxes in PN, MOS1 and MOS2.
Finally, in case of no detection or detection with significant errors (less than $1\sigma $ detections), we provide an 
upper limit for the flux in the hard band (2--10 keV).
Such sources have the flag {\tt LX2\_10\_UPPER} set to -1.

There are 2,311 matches with the 2XMMi-DR3 catalog.

%
\section{Description of the DR10Q catalog}
\label{s:Catalog_description}
The DR10Q catalog is available as a binary FITS table file at the
SDSS public website\footnote{http://www.sdss3.org/dr10/algorithms/qso\_catalog.php}.
The FITS header contains all of the required documentation
(format, name, unit of each column).  
\Tab{CatalogFormat} provides a summary of the information
contained in each of the columns in the catalog.

\par\noindent
Notes on the catalog columns:\\
\smallskip
\noindent
1. The DR10 object designation, given by the format \hbox{SDSS Jhhmmss.ss+ddmmss.s}; only the final~18
characters  are listed in the catalog (i.e., the character string \hbox{"SDSS J"}  is dropped).
The coordinates in the object name follow IAU convention and are truncated, not rounded.

\noindent
2-3. The J2000 coordinates (Right Ascension and Declination) in decimal degrees.  
The astrometry is from SDSS-DR10 \citep[see][]{DR9}.

\noindent
4.  The 64-bit integer that uniquely describes the objects
that are listed in the SDSS (photometric and spectroscopic) catalogs ({\tt THING\_ID}).

\noindent
5-7. Information about the spectroscopic observation (Spectroscopic plate number, 
Modified Julian Date, and spectroscopic fiber number) used to
determine the characteristics of the spectrum.
These three numbers are unique for each spectrum, and
can be used to retrieve the digital spectra from the public SDSS database.
When an object has been observed several times, we selected the best quality spectrum as defined by the SDSS pipeline \citep{bolton2012}, i.e. with {\tt SPECPRIMARY}~=~1.

\noindent
8. Redshift from the visual inspection, {\tt Z\_VI}.

\noindent
9. Redshift from the BOSS pipeline \citep{bolton2012}.

\noindent
10. Error on the BOSS pipeline redshift estimate.

\noindent
11. ZWARNING flag from the pipeline. ZWARNING~$>$~0 indicates uncertain results in 
the redshift-fitting code \citep{bolton2012}.

\noindent
12.  Automatic redshift estimate using a linear combination of four principal components \citep[see Section 4 of ][ for details]{paris2012}.
When the velocity difference between the automatic PCA and visual inspection redshift estimates is larger than 
3000~${\rm km \ s^{-1}}$, this PCA redshift is set to $-1$.

\noindent
13. Error on the automatic PCA redshift estimate.
If the PCA redshift is set to $-1$, the associated error is also set to $-1$.

\noindent
14. Estimator of the PCA continuum quality (between 0 and 1) as given in Eq.(11) of \cite{paris2011}.

\noindent
15-17. Redshifts measured from the \CIV , the \CIII\ complex and the \MgII\ emission lines 
from a linear combination of five principal components (see P\^aris et al. 2012). 
The line redshift is estimated using the maximum of each emission line, contrary to {\tt Z\_PCA} (column \#12) that is a global estimate using all the information available in a given spectrum.

\noindent
18. Morphological information: objects classified as point source by the SDSS photometric pipeline have {\tt SDSS\_MORPHO}~=~0 while extended quasars have {\tt SDSS\_MORPHO}~=~1.
The vast majority of the quasars included in the present catalog are unresolved ({\tt SDSS\_MORPHO}~=~0) as it is a requirement of the main quasar target selection \citep{ross2012}. 

\noindent
19-21.
The main target selection information is tracked with the {\tt BOSS\_TARGET1} flag bits \citep[19; see Table 2 in ][ for a full description]{ross2012}.
Ancillary program target selection is tracked 
with the {\tt ANCILLARY\_TARGET1} (20) and {\tt ANCILLARY\_TARGET2} (21) flag bits.
The bit values and the corresponding program names are listed in \cite{dawson2013} and Appendix B. 

\noindent
22. A quasar previously known from the SDSS-DR7 quasar catalog \citep{schneider2010}  has an entry equal to 1, and 0 otherwise.
During Year 1 and 2, all SDSS-DR7 quasars with $z \geq 2.15$ were re-observed.
After Year 2, all SDSS-DR7 quasars with $z \geq 1.8$ have been re-observed.

\noindent
23-25. Spectroscopic plate number, Modified Julian Date, and spectroscopic fiber number
in SDSS-DR7.

\noindent
26. {\tt Uniform }flag. See \Sec{uniform}.

\noindent
27. Spectral index $\alpha _{\nu}$.
The continuum is approximated by a power-law, $f_{cont} \propto \nu ^{\alpha _{\nu}}$, and is computed in emission line free regions: 1,450-1,500 \AA , 1,700-1,850 \AA\ and 1,950-2,750\AA\ in the quasar rest frame.

\noindent
28. Median signal-to-noise ratio per pixel computed over the whole spectrum.

\noindent
29-31.  Median signal-to-noise ratio per pixel computed over the windows 1,650-1,750 \AA\ (29),
2,950-3,050 \AA\ (30), 2,950-3,050 \AA\ (31)
 in the quasar rest frame. If the wavelength range is not covered by the BOSS spectrum, the value is set to $-1$.

\noindent
32-35. FWHM (${\rm km \ s^{-1}}$), blue and red half widths at half-maximum (HWHM;
the sum of the latter two equals FWHM), and amplitude \citep[in units of the median rms pixel noise, see Section~4 of ][]{paris2012} of the \CIV\  emission line. 
If the emission line is not in the spectrum, the red and blue HWHM and the FWHM are set to $-1$.

\noindent
36-37. Rest frame equivalent width and corresponding uncertainty in \AA\ of the \CIV\ emission line. If the emission line is not in the spectrum, these quantities are set to $-1$.

\noindent
38-41. 
Same as 32-35 for the  \ion{C}{iii]} emission complex.
It is well known that \ion{C}{iii]}${\rm \lambda \lambda }$1909 is blended with \ion{Si}{iii]}${\rm \lambda \lambda }$1892
and to a lesser extent with \ion{Al}{iii]}${\rm \lambda \lambda }$1857. We do not attempt
to deblend these lines. Therefore the redshift and red
HFHM derived for this blend correspond to
\ion{C}{iii]}${\rm \lambda \lambda }$1909. The blue HFWM is obviously affected by the blend.

\noindent
42-43. Rest frame equivalent width and corresponding uncertainty in \AA\ of the \CIII\ emission complex. 

\noindent
44-47.  Same as 32-35 for the \MgII\ emission line. 

\noindent
48-49. Rest frame equivalent width and corresponding uncertainty in \AA\ of the \MgII\ emission line. We do not correct for the \ion{Fe}{ii} emission.

\noindent
50. BAL flag from the visual inspection, {\tt BAL\_FLAG\_VI}. 
If a BAL feature was identified in the course of the visual inspection, {\tt BAL\_FLAG\_VI} is set to 1, 0 otherwise.
Note that  BAL quasars are flagged during the visual inspection at any redshift.

\noindent
51-52. Balnicity index \citep[BI; ][]{weymann1991} for \CIV\ troughs, and their errors, expressed in ${\rm km \ s^{-1}}$. See definition in Section 5 of \cite{paris2012}.
The Balnicity index is measured for quasars with $z > 1.57$ only so that the trough enters into the BOSS wavelength region.
If the BAL flag from the visual inspection is set to 1 and the BI is equal to 0, this means either that there is no 
\CIV\ trough (but a trough is seen in another transition) or that the trough seen during the visual inspection does 
not meet the formal requirement of the BAL definition. 
In cases with bad fits to the continuum, the balnicity index and its error are set to $-1$.

\noindent
53-54. Absorption index, and its error, for \CIV\ troughs expressed in ${\rm km \ s^{-1}}$. See definition in Section 5 of \cite{paris2012}.
In cases with bad continuum fit, the absorption index and its error are set to $-1$.

\noindent
55. Following \cite{trump2006}, we calculate the reduced $\chi^2$ which we call $\chi^2 _{{\rm trough}}$ for each \CIV\ trough
from Eq. (3) in \cite{paris2012}.
We require that troughs have $\chi^2_{\rm trough}$~$>$~10 to be considered as
true troughs.

\noindent
56. Number of \CIV\ troughs of width larger than 2,000~km~s$^{-1}$.

\noindent
57-58. Limits of the velocity range in which 
\CIV\ troughs of width larger than 2,000~km~s$^{-1}$ 
and reaching at least 10\% below the continuum 
are to be found.  Velocities are positive bluewards and the zero of the
scale is at {\tt Z\_VI}.
So if there are multiple troughs, this goes from one end of the first one  to the other end of the last one.

\noindent 
59. Number of troughs of width larger than 450~km~s$^{-1}$.

\noindent
60-61. Same as 59-60 for \CIV\ troughs of width larger than 450~km~s$^{-1}$.

\noindent
62-64. Rest frame equivalent width in \AA\ of \SiIV, \CIV\ and \AlIII\  troughs detected in BAL quasars
with BI $>$ 500 ${\rm km \ s^ {-1}}$ and SNR\_1700 $>$~5.
They are set to 0 otherwise or in cases where no trough is detected and to $-1$ if the continuum is not reliable.

\noindent
65-66. The SDSS Imaging Run number and the Modified Julian Date (MJD) of the
photometric observation used in the catalog.

\noindent
67-70. Additional SDSS processing information: the
photometric processing rerun number; the camera column (1--6) containing
the image of the object, the field number of the run containing the object,
and the object identification number
\citep[see][for descriptions of these parameters]{stoughton2002}.

\noindent
71-72. DR10 PSF fluxes, expressed in nanomaggies, and inverse variances (not corrected for Galactic extinction) in the five SDSS filters. 

\noindent
73-74. DR10 PSF AB magnitudes and errors (not corrected for Galactic extinction) in the five SDSS filters \citep{lupton1999}.

\noindent
75. DR8 PSF fluxes (not corrected for Galactic extinction), expressed in nanommagies, in the five SDSS filters. This photometry is the one that was used for the main quasar target selection \citep{ross2012}.

\noindent
76. The absolute magnitude in the $i$ band at $z=2$ calculated after correction for
Galactic extinction and assuming
\hbox{$H_0$ = 70 km s$^{-1}$ Mpc$^{-1}$,}
$\Omega_{\rm M}$~=~0.3, $\Omega_{\Lambda}$~=~0.7, and a power-law (frequency)
continuum index of~$-0.5$.
The K-correction is computed using Table~4 from \cite{richards2006}.
We use the SDSS primary photometry to compute this value.

\noindent
77. The $\Delta (g-i)$ color, which is the difference in the Galactic
extinction corrected $(g-i)$ for the quasar and that of the mean of the
quasars at that redshift .  If $\Delta (g-i)$ is not defined for the quasar,
which occurs for objects at either \hbox{$z < 0.12$} or \hbox{$z > 5.12$};
the column will contain~``$-999.999$".

\noindent
78. Galactic extinction in the five SDSS bands based on the maps of
\cite{schlegel1998}. The quasar target selection was done using these maps.

\noindent
79. Galactic extinction in the five SDSS bands based on \cite{schlafly2011}.

\noindent
80. The logarithm of the Galactic neutral hydrogen column density along the
line of sight to the quasar. These values were
estimated via interpolation of the 21-cm data from \cite{stark1992},
using the COLDEN software provided by the {\it Chandra} X-ray Center.
Errors associated with the interpolation are expected to
be typically less than $\approx 1\times 10^{20}$~cm$^{-2}$ 
\citep[e.g., see \S5 of][]{elvis1994b}.

\noindent
81. The logarithm of the vignetting-corrected count rate (photons s$^{-1}$)
in the broad energy band \hbox{(0.1--2.4 keV)} from the
{\it ROSAT} All-Sky Survey Faint Source Catalog \citep{voges2000} and the
{\it ROSAT} All-Sky Survey Bright Source Catalog \citep{voges1999}.
The matching radius was set to 30$''$ (see \Sec{ROSAT}).

\noindent
82. The S/N of the {\it ROSAT} measurement.

\noindent
83. Angular separation between the SDSS and {\it ROSAT} All-Sky Survey
locations (in arcseconds).

\noindent
84. Number of XMM-Newton matches in a 5" radius around the SDSS-DR10 quasar positions.

\noindent
85. Soft X-ray flux (0.5--2 keV) from XMM-Newton matching, expressed in ${\rm erg \ cm^{-2} \ s^{-1}}$. In case of several observations, the one with the longest exposure time is used to compute the flux.

\noindent
86. Hard X-ray flux (2--12 keV) from XMM-Newton matching, expressed in ${\rm erg \ cm^{-2} \ s^{-1}}$.

\noindent
87. X-ray luminosity in the 0.5--2 keV band of XMM-Newton, expressed in ${\rm erg \ s^{-1}}$. This value is computed using the visual inspection redshift ({\tt Z\_VI}) and ${\rm H_0 = 70 \ km s^{-1} Mpc^{-1}, \ \Omega _m = 0.3, \ \Omega _{\Lambda} = 0.7}$.

\noindent
88. X-ray luminosity in the 2--12 keV band of XMM-Newton, expressed in ${\rm erg \ s^{-1}}$. This value is computed using the visual inspection redshift ({\tt Z\_VI}) and ${\rm H_0 = 70 \ km s^{-1} Mpc^{-1}, \ \Omega _m = 0.3, \ \Omega _{\Lambda} = 0.7}$.

\noindent
89. In case of unreliable or no detection 
in the 2--10 keV band the flux reported in column \#86 is an upper limit. 
In that case, the {\tt LUMX2\_10\_UPPER} flag listed in this column is set to $-1$.

\noindent
90. Angular separation between the XMM-Newton and SDSS-DR10 locations, expressed in arcsec.

\noindent
91. If a SDSS-DR10 quasar matches with GALEX photometring, {\tt GALEX\_MATCHED} is set to 1, 0 if not.

\noindent
92-95. UV fluxes and inverse variances from GALEX, aperture-photometered from the original GALEX images in the 
two bands FUV and NUV. The fluxes are expressed in nanomaggies.

\noindent
96-97. The $J$ magnitude and error from the Two Micron All Sky Survey
All-Sky Data Release Point Source Catalog \citep{cutri2003} using
a matching radius of ~2.0$''$ (see \Sec{2MASS}).  A non-detection by 2MASS is indicated by a "0.000" in these columns.  Note that the 2MASS measurements are Vega-based, not AB, magnitudes.  

\noindent
98-99. S/N in the $J$ band and corresponding 2MASS jr\_d flag that gives the meaning of the peculiar values of the magnitude and its error\footnote{see http://www.ipac.caltech.edu/2mass/releases/allsky/doc/ explsup.html}.

\noindent
100-103. Same as 96-99 for the $H$-band.

\noindent
104-107. Same as 96-99 for the $K$-band.

\noindent
108. Angular separation between the SDSS and 2MASS positions (in arcseconds).

\noindent
109-110. The $w1$  magnitude and error from the Wide-field Infrared Survey Explorer
\citep[WISE;][]{wright2010} All-Sky Data Release Point Source Catalog  using a matching radius of 2$"$.

\noindent
111-112 S/N and $\chi^2$ in the WISE $w1$ band.

\noindent
113-116. Same as 109-112 for the $w2$-band.

\noindent
117-120. Same as 109-112 for the $w3$-band.

\noindent
121-124. Same as 109-112 for the $w4$-band.

\noindent
125. WISE contamination and confusion flag.

\noindent
126. Angular separation between SDSS and WISE positions (in arcsec).

\noindent
127. If a SDSS-DR10 quasar matches UKIDSS aperture-photometering data, {\tt UKIDSS\_MATCHED} is set to 1, it is set to 0 if not.

\noindent
128-135. Flux and error from UKIDSS, aperture-photometered from the original UKIDSS images in the four bands Y, J, 
H and K. The fluxes and errors are expressed in ${\rm W \ m^{-2} \ Hz^{-1}}$.

\noindent
136. If there is a source in the FIRST radio catalog (version February 2012) within $2.0"$
of the quasar position, the {\tt FIRST\_MATCHED} flag provided in this column is set to 1, O if not. If the quasar lies outside of the FIRST footprint, it is set to $-1$.

\noindent
137. This column contains the FIRST peak flux density, expressed in mJy.

\noindent
138. The S/N of the FIRST source whose flux is given in column~136.

\noindent
139. Angular separation between the SDSS and FIRST positions (in arcsec).

\addtocounter{table}{1}

%
\section{Conclusions}
\label{s:Summary}

We have presented the second quasar catalog of the SDSS-III/BOSS survey based on the first three years of observations.
It contains 166,583 quasars, 117,668 having $z>2.15$, with robust identification from visual inspection and refined redshift 
measurements based on the results of a principal component analysis of the spectra. 
The present catalog is almost twice as large as the catalog of the previous release \citep[DR9Q;][]{paris2012}
As part of the DR10Q catalog, we also release a catalog of 16,461 BAL quasars and their properties
together with a multi-wavelength matching 
with GALEX, 2MASS, UKIDSS, WISE, FIRST, RASS and XMM observations.

The next SDSS public release, SDSS-DR12, is scheduled for December 2014 and will contain more than 250,000 quasars, including about 200,000 $z>2.15$ quasars.

\begin{acknowledgements}
	  I.P. received partial support from Center of Excellence in Astrophysics and Associated Technologies (PFB 06).
      The French Participation Group to SDSS-III was supported by the Agence Nationale de la Recherche under contracts ANR-08-BLAN-0222
and ANR-12-BS05-0015.  
	  A.D.M. is a research fellow of the Alexander von Humboldt Foundation of Germany and was partially supported through NSF Grant 1211112 and NASA ADAP award NNX12AE38G. 
	    
Funding for SDSS-III has been provided by the Alfred P. Sloan Foundation, the Participating Institutions, the National Science Foundation, and the U.S. Department of Energy Office of Science. The SDSS-III web site is http://www.sdss3.org/.

SDSS-III is managed by the Astrophysical Research Consortium for the Participating Institutions of the SDSS-III Collaboration including the University of Arizona, the Brazilian Participation Group, Brookhaven National Laboratory, Carnegie Mellon University, University of Florida, the French Participation Group, the German Participation Group, Harvard University, the Instituto de Astrofisica de Canarias, the Michigan State/Notre Dame/JINA Participation Group, Johns Hopkins University, Lawrence Berkeley National Laboratory, Max Planck Institute for Astrophysics, Max Planck Institute for Extraterrestrial Physics, New Mexico State University, New York University, Ohio State University, Pennsylvania State University, University of Portsmouth, Princeton University, the Spanish Participation Group, University of Tokyo, University of Utah, Vanderbilt University, University of Virginia, University of Washington, and Yale University.

\end{acknowledgements}

\bibliographystyle{aa}
\bibliography{DR10Q}

\longtab{3}{
\begin{longtable}{clcl}
\caption{\label{t:CatalogFormat} DR10Q catalog format}\\
\hline\hline
Column & Name & Format & Description$^a$ \\
\hline
\endfirsthead
\caption{continued.}\\
\hline\hline
Column & Name  &  Format & Description \\
\hline
\endhead
\hline
\endfoot
%
1    & SDSS\_NAME                      &  STRING      & SDSS-DR10 designation  hhmmss.ss+ddmmss.s (J2000)\\
2    & RA                              &  DOUBLE   & Right Ascension in decimal degrees (J2000)\\
3    & DEC                             & DOUBLE    & Declination in decimal degrees (J2000)\\
4    & THING\_ID                       &  INT32       & Thing\_ID \\
5    & PLATE                           & INT32         & Spectroscopic Plate number \\
6    & MJD                             & INT32         & Spectroscopic MJD \\
7    & FIBERID                         & INT32         & Spectroscopic Fiber number \\
%
\hline
%
%
8    & Z\_VI                           &  DOUBLE    & Redshift from visual inspection \\
9    & Z\_PIPE                        &  DOUBLE   & Redshift from BOSS pipeline \\
10    & ERR\_ZPIPE                     &  DOUBLE     & Error on BOSS pipeline redshift \\
11   & ZWARNING                        & INT32          & ZWARNING flag  \\
12  & Z\_PCA                           & DOUBLE      & Refined PCA redshift \\
13  & ERR\_ZPCA                        & DOUBLE       & Error on refined PCA redshift \\
14  & PCA\_QUAL                        & DOUBLE      & Estimator of the PCA continuum quality\\
15   & Z\_CIV                          & DOUBLE       & Redshift of \CIV\ emission line \\
16   & Z\_CIII                         & DOUBLE       & Redshift of \CIII\ emission complex \\
17   & Z\_MGII                         & DOUBLE       & Redshift of \MgII\ emission line\\
\hline
%
%
18   & SDSS\_MORPHO                 &   INT32     & SDSS morphology flag 0 = point source 1 = extended \\
19   & BOSS\_TARGET1                 &  INT64    & BOSS target flag for main survey  \\
20   & ANCILLARY\_TARGET1       &   INT64   & BOSS target flag for ancillary programs \\
21   & ANCILLARY\_TARGET2       &   INT64   & BOSS target flag for ancillary programs  \\
22   &SDSS\_DR7                           & INT32      & 1 if the quasar is known from DR7\\ 
23  & PLATE\_DR7                         & INT32       & SDSS-DR7 spectroscopic Plate number if the quasar is known from DR7 \\
24  & MJD\_DR7                           &  INT32       & SDSS-DR7 spectroscopic MJD  if the quasar is known from DR7\\
25  & FIBERID\_DR7                     & INT32       & SDSS-DR7 spectroscopic Fiber number  if the quasar is known from DR7\\
26   & UNIFORM                           & INT32        & Uniform sample flag \\
27   & ALPHA\_NU                       &  FLOAT    & Spectral index measurement $\alpha _{\nu}$ \\
\hline
%
%
28   & SNR\_SPEC                      &    FLOAT    & Median signal-to-noise ratio over the whole spectrum\\
29   & SNR\_1700                      &    FLOAT    & Median signal-to-noise ratio in the window 1,650 - 1,750\AA\ (rest frame)\\
30   & SNR\_3000                      &    FLOAT    & Median signal-to-noise ratio in the window 2,950 - 3,050\AA\ (rest frame)\\
31   & SNR\_5150                      &    FLOAT    & Median signal-to-noise ratio in the window 5,100 - 5,250\AA\ (rest frame)\\
32   & FWHM\_CIV                     &   DOUBLE     & FWHM of \CIV\ emission line in ${\rm km \ s^{-1}}$ \\
33   & BHWHM\_CIV                  &    DOUBLE     & Blue HWHM of \CIV\ emission line in ${\rm km \ s^{-1}}$ \\
34   & RHWHM\_CIV                  &    DOUBLE   & Red HWHM of \CIV\ emission line in ${\rm km \ s^{-1}}$ \\
35   & AMP\_CIV                        &   DOUBLE      & Amplitude of \CIV\ emission line in units of median rms pixel noise \\
36   & REWE\_CIV                     &   DOUBLE     & Rest frame equivalent width of \CIV\ emission line in \AA \\
37   & ERR\_REWE\_CIV           &  DOUBLE     & Uncertainty on the rest frame equivalent width of \CIV\ emission line in \AA \\
38   & FWHM\_CIII                    &     DOUBLE       & FWHM of \CIII\ emission complex in ${\rm km \ s^{-1}}$ \\
39   & BHWHM\_CIII                 &     DOUBLE       & Blue HWHM of \CIII\ emission line in ${\rm km \ s^{-1}}$ \\
40   & RHWHM\_CIII                 &     DOUBLE      & Red HWHM of \CIII\ emission line in ${\rm km \ s^{-1}}$ \\
41   & AMP\_CIII                       &    DOUBLE      & Amplitude of \CIII\ emission complex in units of median rms pixel noise\\
42   & REWE\_CIII                    &      DOUBLE     & Rest frame equivalent width of \CIII\ emission line in \AA \\
43  & ERR\_REWE\_CIII           &  DOUBLE     & Uncertainty on the rest frame equivalent width of \CIII\ emission complex in \AA \\
44   & FWHM\_MGII                  &    DOUBLE       & FWHM of \MgII\ emission line in ${\rm km \ s^{-1}}$ \\
45   & BHWHM\_MGII               &      DOUBLE      & Blue HWHM of \MgII\ emission line in ${\rm km \ s^{-1}}$ \\
46   & RHWHM\_MGII               &      DOUBLE     & Red HWHM of \MgII\ emission line in ${\rm km \ s^{-1}}$ \\
47   & AMP\_MGII                     &     DOUBLE      & Amplitude of \MgII\ emission line in units of median rms pixel noise \\
48   & REWE\_MGII                  &      DOUBLE     & Rest frame equivalent width of \MgII\ emission line in \AA \\
49  & ERR\_REWE\_MGII         &  DOUBLE     & Uncertainty on the rest frame equivalent width of \MgII\ emission in \AA \\
\hline
50   & BAL\_FLAG\_VI                  &       INT32       & BAL flag from visual inspection \\
51   & BI\_CIV                              &     DOUBLE       & Balnicity index of \CIV\ trough (in ${\rm km \ s^{-1}}$) \\
52   & ERR\_BI\_CIV                    &     DOUBLE      & Error on the Balnicity index of \CIV\ trough (in ${\rm km \ s^{-1}}$) \\
53   & AI\_CIV                             &   DOUBLE       & Absorption index of \CIV\ trough (in ${\rm km \ s^{-1}}$) \\
54   & ERR\_AI\_CIV                   &    DOUBLE       & Error on the absorption index of \CIV\ trough (in ${\rm km \ s^{-1}}$) \\
55   & CHI2THROUGH                 &    DOUBLE       & $\chi^2$ of the trough from Eq. (3) in \cite{paris2012}\\
56   & NCIV\_2000                       &  INT32        & Number of distinct \CIV\ troughs of width larger than 2,000~km~s$^{-1}$ \\
57   & VMIN\_CIV\_2000              &    DOUBLE    & Minimum velocity of the \CIV\ troughs defined in row 53 (${\rm km \ s^{-1}}$) \\
58   & VMAX\_CIV\_2000             &     DOUBLE    & Maximum velocity of the \CIV\ troughs defined in row 53  (in ${\rm km \ s^{-1}}$) \\
59   & NCIV\_450                         &  INT32        & Number of distinct \CIV\ troughs of width larger than 450~km~s$^{-1}$ \\
60   & VMIN\_CIV\_450               &    DOUBLE     & Minimum velocity of the \CIV\ troughs defined in row 55 (in ${\rm km \ s^{-1}}$) \\
61   & VMAX\_CIV\_450              &    DOUBLE     & Maximum velocity of the \CIV\ troughs defined in row 55 (in ${\rm km \ s^{-1}}$) \\
62   & REW\_SIIV                       &  DOUBLE        & rest frame equivalent width of the \SiIV\ trough \\
63   & REW\_CIV                         &  DOUBLE      & rest frame equivalent width of the \CIV\ trough  \\
64   & REW\_ALIII                      & DOUBLE          & rest frame equivalent width of the Al~{\sc iii} trough \\
\hline
%
%
65   & RUN\_NUMBER                     & INT32      & SDSS Imaging Run Number of photometric measurements \\
66   & PHOTO\_MJD                       & INT32     & Modified Julian Date of imaging observation \\
67   & RERUN\_NUMBER                 & STRING    & SDSS Photometric Processing Rerun Number \\
68   & COL\_NUMBER                      & INT32      & SDSS Camera Column Number (1-6) \\
69  & FIELD\_NUMBER                   & INT32      & SDSS Field Number \\
70   & OBJ\_ID                                & STRING      & SDSS Object Identification Number \\
%
%
71   & PSFFLUX                             &      FLOAT[5]       & flux in the $u$,$g$,$r$,$i$ and $z$-bands (not corrected for Galactic extinction)\\
72   & IVAR\_PSFLUX                    &       FLOAT[5]      & inverse variance of $u$,$g$,$r$,$i$ and $z$ fluxes \\
73   & PSFMAG                          & FLOAT[5]            & PSF magnitudes in $u$,$g$,$r$,$i$ and $z$-bands (not corrected for Galactic extinction)\\
74   & ERR\_PSFMAG                          & FLOAT[5]            & error in $u$,$g$,$r$,$i$ and $z$ PSF magnitudes\\
75   & TARGET\_FLUX       & FLOAT[5]       & TARGET flux in the $u$,$g$,$r$,$i$ and $z$-bands (not corrected for Galactic extinction)\\
76   & MI                                      &  FLOAT   & $M_{\rm i}\left[{\rm z = 2} \right] \left( H_0 = 67.8 {\rm km \ s^{-1} \ Mpc^{-1}}, \ \Omega _M = 0.308, \ \Omega _{\Lambda} = 0.692, \ \alpha _{\nu} = -0.5 \right)$ \\
77   & DGMI                                 &  FLOAT   & $\Delta (g-i) = (g-i) - \langle (g-i) \rangle _{\rm redshift}$ (Galactic extinction corrected) \\
78   & GAL\_EXT                           &    FLOAT[5]       & Galactic extinction in the 5 SDSS bands from \cite{schlegel1998} \\
79   & GAL\_EXT\_RECAL             & FLOAT[5] & Galactic extinction in the 5 SDSS bands from \cite{schlafly2011}\\
80   & HI\_GAL                          &    FLOAT       & ${\rm log} N_{\rm H}$ (logarithm of Galactic \HI\ column density in ${\rm cm^{-2}}$)\\
\hline
%
%
81   & RASS\_COUNTS             &     DOUBLE      & log RASS full band count rate (counts s$^{-1}$)\\
82   & RASS\_COUNTS\_SNR   &      FLOAT      & S/N of the RASS count rate \\
83   & SDSS2ROSAT\_SEP        &      DOUBLE      & SDSS-RASS separation in arcsec \\
84   & N\_DETECTION\_XMM          & INT32  & Number of detections in XMM-Newton \\
85   & FLUX05\_2KEV               & DOUBLE & Soft (0.5-2 keV) X-ray flux from XMM-Newton (${\rm erg \ cm^{-2} \ s^{-1}}$) \\
86   & FLUX2\_12KEV               & DOUBLE & Hard (2-12 keV) X-ray flux from XMM-Newton (${\rm erg \ cm^{-2} \ s^{-1}}$) \\
87   & LUM05\_2KEV               & DOUBLE & Soft (0.5-2 keV) X-ray luminosity from XMM-Newton (${\rm erg \ s^{-1}}$) using Z\_VI \\
     &                           &        & and $H_0 = 70 {\rm km \ s^{-1} \ Mpc^{-1}}, \ \Omega _M = 0.3, \ \Omega _{\Lambda} = 0.7$\\
88  & LUM2\_12KEV               & DOUBLE & Hard (2-12 keV) X-ray luminosity from XMM-Newton (${\rm erg \ s^{-1}}$) using Z\_VI\\
    &                           &        &  and $H_0 = 70 {\rm km \ s^{-1} \ Mpc^{-1}}, \ \Omega _M = 0.3, \ \Omega _{\Lambda} = 0.7$\\
89  & LUMX2\_10\_UPPER & DOUBLE & Flag for upper limit of hard X-ray flux (in col. \#86) \\
90  & SDSS2XMM\_SEP & DOUBLE & SDSS-XMM-Newton separation in arcsec \\
91  & GALEX\_MATCHED            & SHORT  & GALEX match \\
92   & FUV                       &    DOUBLE       & $fuv$ flux (GALEX) \\
93   & FUV\_IVAR             &      DOUBLE     & Inverse variance of $fuv$ flux \\
94   & NUV                       &    DOUBLE       & $nuv$ flux (GALEX) \\
95   & NUV\_IVAR              &     DOUBLE      & Inverse variance of $nuv$ flux \\
%
96   & JMAG                             &    DOUBLE       & $J$ magnitude (Vega, 2MASS) \\
97   & ERR\_JMAG                   &   DOUBLE      & Error in $J$ magnitude \\
98   & JSNR                              &    DOUBLE       & J-band S/N \\
99   & JRDFLAG                       &    INT32     & J-band photometry flag\\ 
100 & HMAG                            &   DOUBLE  & $H$ magnitude (Vega, 2MASS) \\
101  & ERR\_HMAG                   &     DOUBLE    & Error in $H$ magnitude \\
102  & HSNR                             &   DOUBLE       & H-band S/N \\
103  & HRDFLAG                       &     INT32      & H-band photometry flag\\ 
104  & KMAG                            &  DOUBLE   & $K$ magnitude (Vega, 2MASS) \\
105  & ERR\_KMAG                  &     DOUBLE      & Error in $K$ magnitude \\
106  & KSNR                              &   DOUBLE       & K-band S/N \\
107  & KRDFLAG                       &      INT32      & K-band photometry flag\\ 
108  & SDSS2MASS\_SEP         &   DOUBLE  & SDSS-2MASS separation in arcsec \\
109  & W1MAG                         &     DOUBLE      & $w1$ magnitude (Vega, WISE) \\
110  & ERR\_W1MAG               &     DOUBLE      & Error in $w1$ magnitude \\
111  & W1SNR                         &   DOUBLE        & S/N in w1 band    \\
112  & W1CHI2                         &   DOUBLE       & $\chi^2$ in w1 band \\
113  & W2MAG                        &      DOUBLE    & $w2$ magnitude (Vega, WISE) \\
114  & ERR\_W2MAG              &     DOUBLE     & Error in $w2$ magnitude \\
115  & W2SNR                         &   DOUBLE        & S/N in w1 band    \\
116  & W2CHI2                         &   DOUBLE       & $\chi^2$ in w1 band \\
117  & W3MAG                        &      DOUBLE    & $w3$ magnitude (Vega, WISE) \\
118  & ERR\_W3MAG              &     DOUBLE       & Error in $w3$ magnitude \\
119  & W3SNR                         &   DOUBLE        & S/N in w1 band    \\
120  & W3CHI2                         &   DOUBLE       & $\chi^2$ in w1 band \\
121  & W4MAG                        &      DOUBLE    & $w4$ magnitude  (Vega, WISE) \\
122  & ERR\_W4MAG               &      DOUBLE   & Error in $w4$ magnitude \\
123  & W4SNR                         &   DOUBLE       & S/N in w1 band    \\
124  & W4CHI2                         &   DOUBLE       & $\chi^2$ in w1 band \\
125  & CC\_FLAGS                 & STRING & WISE contamination and confusion flag \\
126  & SDSS2WISE\_SEP             &      DOUBLE     & SDSS-WISE separation in arcsec \\
127  & UKIDSS\_MATCHED           & SHORT & UKIDSS Matched \\
128  & YFLUX    & DOUBLE & Y-band flux from UKIDSS (in ${\rm W \ m^{-2} \ Hz^{-1}}$) \\
129  & YFLUX\_ERR & DOUBLE & Error in Y-band flux from UKIDSS (in ${\rm W \ m^{-2} \ Hz^{-1}}$) \\
130  & JFLUX    & DOUBLE & J-band flux from UKIDSS (in ${\rm W \ m^{-2} \ Hz^{-1}}$)\\
131  & JFLUX\_ERR & DOUBLE & Error in J-band flux from UKIDSS (in ${\rm W \ m^{-2} \ Hz^{-1}}$) \\
132  & HFLUX    & DOUBLE & H-band flux from UKIDSS (in ${\rm W \ m^{-2} \ Hz^{-1}}$)\\
133  & HFLUX\_ERR & DOUBLE & Error in H-band flux from UKIDSS (in ${\rm W \ m^{-2} \ Hz^{-1}}$)\\
134  & KFLUX    & DOUBLE & K-band flux from UKIDSS (in ${\rm W \ m^{-2} \ Hz^{-1}}$)\\
135  & KFLUX\_ERR & DOUBLE & Error in K-band flux from UKIDSS  (in ${\rm W \ m^{-2} \ Hz^{-1}}$) \\
136  & FIRST\_MATCHED & INT & FIRST matched \\
137  & FIRST\_FLUX                     &   DOUBLE      & FIRST peak flux density at 20 cm expressed in mJy \\
138  & FIRST\_SNR                      &    DOUBLE     & S/N of the FIRST flux density \\
139  & SDSS2FIRST\_SEP              &       DOUBLE  & SDSS-FIRST separation in arcsec \\
\hline
\multicolumn{4}{l}{$^a$ All magnitudes are PSF magnitudes}
\end{longtable}
}

\begin{appendix}
\section{List of target selection flags}
\label{ap:QTSflag}

\Tab{nbQTS} presents the list of target selection bits used to select the parent sample of quasar targets (i.e., without serendipitous quasars from galaxy targets) together with the result of the visual inspection.

The main quasar target selection \citep{ross2012} is encoded with the target selection flag {\tt BOSS\_TARGET1} while the various ancillary program target selections are encoded via the flags {\tt ANCILLARY\_TARGET1} and {\tt ANCILLARY\_TARGET2} \citep{dawson2013}. A given quasar target can be selected by various algorithms or different ancillary programs, and hence the numbers given in \Tab{nbQTS} are not cumulative.

\begin{table*}
\centering                          
\begin{tabular}{l r r  r r r r r r}
\hline                      
\hline                
Selection      & Maskbits    & \# Objects   & \# QSO     & \# QSO $z>2.15$ & \# STAR    &  \# GALAXY   & \# ?     & \# BAD\\
\hline
                    & BOSS\_TARGET1 & & & & & & & \\
\hline
       QSO\_CORE  &   10 &     3402 &     1349 &     1089 &     1941 &       64 &     40 &      8 \\
      QSO\_BONUS  &   11 &     4149 &      788 &      469 &     3231 &       84 &     35 &     11 \\
 QSO\_KNOWN\_MIDZ &   12 &    15020 &    14919 &    14781 &       34 &        1 &     40 &     26 \\
QSO\_KNOWN\_LOHIZ &   13 &       57 &       57 &        1 &        0 &        0 &      0 &      0 \\
         QSO\_NN  &   14 &   139310 &    92311 &    74711 &    44192 &     1119 &    781 &    907 \\
     QSO\_UKIDSS  &   15 &       48 &       27 &       22 &       19 &        2 &      0 &      0 \\
 QSO\_LIKE\_COADD &   16 &     1368 &      314 &      229 &      918 &       55 &     61 &     20 \\
       QSO\_LIKE  &   17 &   109433 &    66565 &    46193 &    38749 &     2023 &   1127 &    969 \\
 QSO\_FIRST\_BOSS &   18 &     6195 &     4884 &     3314 &      638 &      240 &    359 &     74 \\
        QSO\_KDE  &   19 &   162988 &    94868 &    73075 &    64277 &     1680 &   1115 &   1048 \\
  QSO\_CORE\_MAIN &   40 &   130024 &    82672 &    67642 &    43885 &     1374 &   1051 &   1042 \\
 QSO\_BONUS\_MAIN &   41 &   264278 &   144539 &   108756 &   108993 &     5473 &   2812 &   2461 \\
    QSO\_CORE\_ED &   42 &    32659 &    21863 &    18553 &     9930 &      269 &    285 &    312 \\
  QSO\_CORE\_LIKE &   43 &    34919 &    25924 &    18945 &     8003 &      446 &    266 &    280 \\
QSO\_KNOWN\_SUPPZ &   44 &       57 &       57 &        1 &        0 &        0 &      0 &      0 \\
\hline
                    & ANCILLARY\_TARGET1 & & & & & & & \\
\hline
       BLAZGVAR   &    6 &        2 &        1 &        0 &        0 &        0 &      1 &      0 \\
          BLAZR   &    7 &        6 &        2 &        0 &        0 &        4 &      0 &      0 \\
         BLAZXR   &    8 &      361 &      102 &        5 &       27 &      202 &     30 &      0 \\
      BLAZXRSAL   &    9 &        3 &        3 &        1 &        0 &        0 &      0 &      0 \\
      BLAZXRVAR   &   10 &        1 &        0 &        0 &        0 &        0 &      1 &      0 \\
      XMMBRIGHT   &   11 &      413 &      314 &       15 &       14 &       84 &      1 &      0 \\
        XMMGRIZ   &   12 &       62 &       14 &        8 &       21 &       18 &      8 &      1 \\
          XMMHR   &   13 &      476 &      141 &       14 &       24 &      294 &     14 &      3 \\
         XMMRED   &   14 &      337 &       57 &        2 &       42 &      234 &      1 &      3 \\
        FBQSBAL   &   15 &        6 &        5 &        2 &        0 &        0 &      1 &      0 \\
        LBQSBAL   &   16 &        6 &        6 &        0 &        0 &        0 &      0 &      0 \\
         ODDBAL   &   17 &       21 &       15 &        6 &        1 &        1 &      4 &      0 \\
          OTBAL   &   18 &       11 &        2 &        0 &        0 &        0 &      9 &      0 \\
        PREVBAL   &   19 &       10 &       10 &        2 &        0 &        0 &      0 &      0 \\
         VARBAL   &   20 &     1007 &      999 &      370 &        0 &        0 &      4 &      4 \\
        QSO\_AAL  &   22 &      324 &      319 &        2 &        1 &        1 &      1 &      2 \\
       QSO\_AALS  &   23 &      622 &      617 &       35 &        0 &        0 &      1 &      4 \\
        QSO\_IAL  &   24 &      195 &      194 &        2 &        0 &        0 &      1 &      0 \\
      QSO\_RADIO  &   25 &      157 &      155 &        8 &        1 &        0 &      1 &      0 \\
  QSO\_RADIO\_AAL &   26 &      104 &      104 &        0 &        0 &        0 &      0 &      0 \\
  QSO\_RADIO\_IAL &   27 &       56 &       55 &        0 &        0 &        0 &      1 &      0 \\
     QSO\_NOAALS  &   28 &       53 &       53 &        1 &        0 &        0 &      0 &      0 \\
        QSO\_GRI  &   29 &     1503 &      584 &      561 &      466 &      300 &    114 &     39 \\
        QSO\_HIZ  &   30 &      393 &        0 &        0 &      351 &        3 &     15 &     24 \\
        QSO\_RIZ  &   31 &     1089 &       70 &       66 &      838 &      111 &     50 &     20 \\
     BLAZGRFLAT   &   50 &       74 &       42 &        6 &        9 &        5 &     15 &      3 \\
      BLAZGRQSO   &   51 &       94 &       58 &       14 &       14 &        3 &     19 &      0 \\
         BLAZGX   &   52 &        8 &        2 &        0 &        5 &        1 &      0 &      0 \\
      BLAZGXQSO   &   53 &       31 &       29 &        2 &        0 &        1 &      1 &      0 \\
        BLAZGXR   &   54 &      100 &       33 &        3 &       12 &       21 &     34 &      0 \\
      CXOBRIGHT   &   58 &       99 &       66 &        2 &        3 &       27 &      3 &      0 \\
         CXORED   &   59 &       17 &        6 &        2 &        3 &        4 &      3 &      1 \\
\hline
                    & ANCILLARY\_TARGET2 & & & & & & & \\
\hline
       HIZQSO82   &    0 &       57 &        2 &        2 &       54 &        1 &      0 &      0 \\
       HIZQSOIR   &    1 &       86 &        2 &        1 &       77 &        0 &      4 &      3 \\
      KQSO\_BOSS  &    2 &      183 &       86 &       40 &       88 &        4 &      4 &      1 \\
        QSO\_VAR  &    3 &     1370 &      865 &      311 &      424 &       79 &      0 &      2 \\
    QSO\_VAR\_FPG &    4 &      579 &      557 &      284 &        5 &        3 &      9 &      5 \\
RADIO\_2LOBE\_QSO &    5 &      806 &      402 &       39 &      265 &       76 &     48 &     15 \\
      QSO\_SUPPZ  &    7 &     2824 &     2805 &       14 &        2 &        0 &      4 &     13 \\
   QSO\_VAR\_SDSS &    8 &    15375 &     5857 &     3003 &     8329 &      224 &    551 &    414 \\
  QSO\_WISE\_SUPP &    9 &     5007 &     2988 &      999 &     1561 &      303 &    135 &     20 \\
\hline                                
\end{tabular}
\caption{
Number of visually inspected DR10 BOSS quasar targets (third column) and identifications in the DR10Q catalog for each target 
selection method (first column; see Table~4 of Ross et al. 2012, and Tables~6 and 7 in the Appendix of Dawson et al. 2013). 
These categories overlap because many objects are selected by multiple algorithms.
}
\label{t:nbQTS}      
\end{table*}

\end{appendix}

\begin{appendix}
\section{Description of the superset of DR10Q}
\label{ap:supersetDR10Q}

We visually inspected 321,729 quasar candidates to produce the DR10Q catalog.
We provide the basic properties (spectroscopic and photometric) of all the quasar targets, together with the result of the visual inspection, in the form of the binary fits file that is available at the SDSS public website\footnote{http://www.sdss3.org/dr10/algorithms/qso\_catalog.php}.
The FITS header contains all of the required documentation
(format, name, unit of each column).

\begin{table*}
\centering                        
\begin{tabular}{c l c l}        
\hline\hline
Column & Name & Format & Description\\
\hline\hline
1    & SDSS\_NAME                      &  STRING      & SDSS-DR10 designation  hhmmss.ss+ddmmss.s (J2000)\\
2    & RA                              &  DOUBLE   & Right Ascension in decimal degrees (J2000)\\
3    & DEC                             & DOUBLE    & Declination in decimal degrees (J2000)\\
4    & THING\_ID                       &  INT32       & Thing\_ID \\
5    & PLATE                           & INT32         & Spectroscopic Plate number \\
6    & MJD                             & INT32         & Spectroscopic MJD \\
7    & FIBERID                         & INT32         & Spectroscopic Fiber number \\
\hline
8    & Z\_VI                           &  DOUBLE    & Redshift from visual inspection \\
9    & Z\_PIPE                        &  DOUBLE   & Redshift from BOSS pipeline \\
10    & ERR\_ZPIPE                     &  DOUBLE     & Error on BOSS pipeline redshift \\
11   & ZWARNING                        & INT32          & ZWARNING flag  \\
12   & CLASS\_PERSON                & INT32       & Classification from the visual inspection \\
13   & Z\_CONF\_PERSON              & INT32       & Redshift confidence from visual inspection \\
\hline
14   & SDSS\_MORPHO                 &   INT32     & SDSS morphology flag 0 = point source 1 = extended \\
15   & BOSS\_TARGET1                 &  INT64    & BOSS target flag for main survey  \\
16   & ANCILLARY\_TARGET1       &   INT64   & BOSS target flag for ancillary programs \\
17   & ANCILLARY\_TARGET2       &   INT64   & BOSS target flag for ancillary programs  \\
\hline
18   & PSFFLUX                             &      FLOAT[5]       & flux in the $u$,$g$,$r$,$i$ and $z$-bands (not corrected for Galactic extinction)\\
19   & IVAR\_PSFLUX                    &       FLOAT[5]      & inverse variance of $u$,$g$,$r$,$i$ and $z$ fluxes \\
20   & PSFMAG                          & FLOAT[5]            & PSF magnitudes in $u$,$g$,$r$,$i$ and $z$-bands (not corrected for Galactic extinction)\\
21   & ERR\_PSFMAG                          & FLOAT[5]            & error in $u$,$g$,$r$,$i$ and $z$ PSF magnitudes\\
22   & TARGET\_FLUX       & FLOAT[5]       & TARGET flux in the $u$,$g$,$r$,$i$ and $z$-bands (not corrected for Galactic extinction)\\
23   & GAL\_EXT                           &    FLOAT[5]       & Galactic extinction in the 5 SDSS bands \citep[from ][]{schlegel1998} \\
\hline
\hline
\end{tabular}
\caption{Description of the file that contains the superset from which we derive the DR10Q catalog. This file contains the result from the visual inspection as described in \Tab{VI_PIPE}.}
\label{t:superset}
\end{table*}

\end{appendix}

\end{document}